\RequirePackage{rotating}
\documentclass[fleqn,usenatbib]{mnras}

\usepackage{newtxtext}
\usepackage[varg,varvw,smallerops]{newtxmath}
\usepackage[utf8]{inputenc}

\usepackage[T1]{fontenc}
\usepackage{ae,aecompl}
\usepackage{xcolor}
\usepackage{multirow}

\usepackage{graphicx}	

\usepackage{amsmath}	
\usepackage{amssymb}	
\usepackage{hyperref}
\usepackage{adjustbox}
\usepackage{rotating}

\usepackage{savesym}
\savesymbol{tablenum}
\usepackage{siunitx}
\restoresymbol{SIX}{tablenum}
\newcounter{ionstage}
\renewcommand{\ion}[2]{\setcounter{ionstage}{#2}%
  \ensuremath{\mathrm{#1\,\scriptstyle\Roman{ionstage}}}}
\newcommand\hii{\ion{H}{2}}
\newcommand\siii{[\ion{S}{3}]}
\newcommand*\chem[1]{\ensuremath{\mathrm{#1}}}

\newcommand\disk{\ensuremath{_{\mathrm{d}}}}

\DeclareSIUnit\msun{\text{M\ensuremath{_\odot}}}
\DeclareSIUnit\mearth{\text{M\ensuremath{_\oplus}}}
\DeclareSIUnit\lsun{\text{L\ensuremath{_\odot}}}
\def\th#1#2{\ensuremath{\theta^{#1}\,\text{Ori~#2}}}

\title[HH~514 in the Orion Nebula]{Photoionized Herbig-Haro objects in the Orion Nebula through deep high-spectral resolution spectroscopy III: HH~514}

\author[J. E. M\'endez-Delgado et al.]
{J. E. M\'endez-Delgado$^{1,2}$ \thanks{E-mail: jemd@iac.es},
C. Esteban$^{1,2}$, J. Garc{\'{\i}}a-Rojas$^{1,2}$ and W. J. Henney$^{3}$  
\\
$^{1}$Instituto de Astrof\'isica de Canarias (IAC), E-38205 La Laguna, Spain\\
$^{2}$Departamento de Astrof\'isica, Universidad de La Laguna, E-38206 La Laguna, Spain\\
$^{3}$Instituto de Radioastronom\'ia y Astrof\'isica, Universidad Nacional Aut\'onoma de M\'exico, Apartado Postal 3-72, 58090 Morelia, Michoac\'an, M\'exico}

\date{Accepted XXX. Received YYY; in original form ZZZ}

\pubyear{2020}
\begin{document}
\label{firstpage}
\pagerange{\pageref{firstpage}--\pageref{lastpage}}
\maketitle

\begin{abstract}
  We analyze the physical conditions and chemical composition of the photoionized
  Herbig-Haro object HH~514,
  which emerges from the proplyd 170-337
  in the core of the Orion Nebula.
  We use high-spectral resolution spectroscopy from UVES
  at the \textit{Very Large Telescope}
  and IFU-spectra from MEGARA at the \textit{Gran Telescopio de Canarias}.
  We observe two components of HH~514, the jet base and a knot,
  with $n_{\rm e}= (2.3 \pm 0.1) \times 10^5 \text{\,cm}^{-3}$
  and $n_{\rm e}= (7 \pm 1) \times 10^4 \text{\,cm}^{-3}$, respectively,
  both with $T_{\rm e}\approx 9000 \text{ K}$.
  We show that the chemical composition of HH~514 is consistent with that of the Orion Nebula,
  except for Fe, Ni and S, which show higher abundances.
  The enhanced abundances of Fe and Ni observed in HH objects compared with the general interstellar medium is usually interpreted as destruction of dust grains.
  The observed sulphur overabundance (more than two times solar) is challenging to explain
  since the proplyd photoevaporation flow from the same disk shows normal sulphur abundance.
  If the aforementioned S-overabundance is due to dust destruction, the formation of sulfides and/or other S-bearing dust reservoirs may be linked to planet formation processes in protoplanetary disks, which filter large sulfide dust grains during the accretion of matter from the disk to the central star.
  We also show that published kinematics of molecular emission close to the central star
  are not consistent with either a disk perpendicular to the optical jet,
  nor with an outflow that is aligned with it.
\end{abstract}


\begin{keywords}
ISM:Abundances – ISM: Herbig–Haro objects – ISM: individual:
Orion Nebula – ISM: individual: HH~514.
\end{keywords}



\section{Introduction}
\label{sec:introduction}

\begin{figure*}
\includegraphics[width=\textwidth]{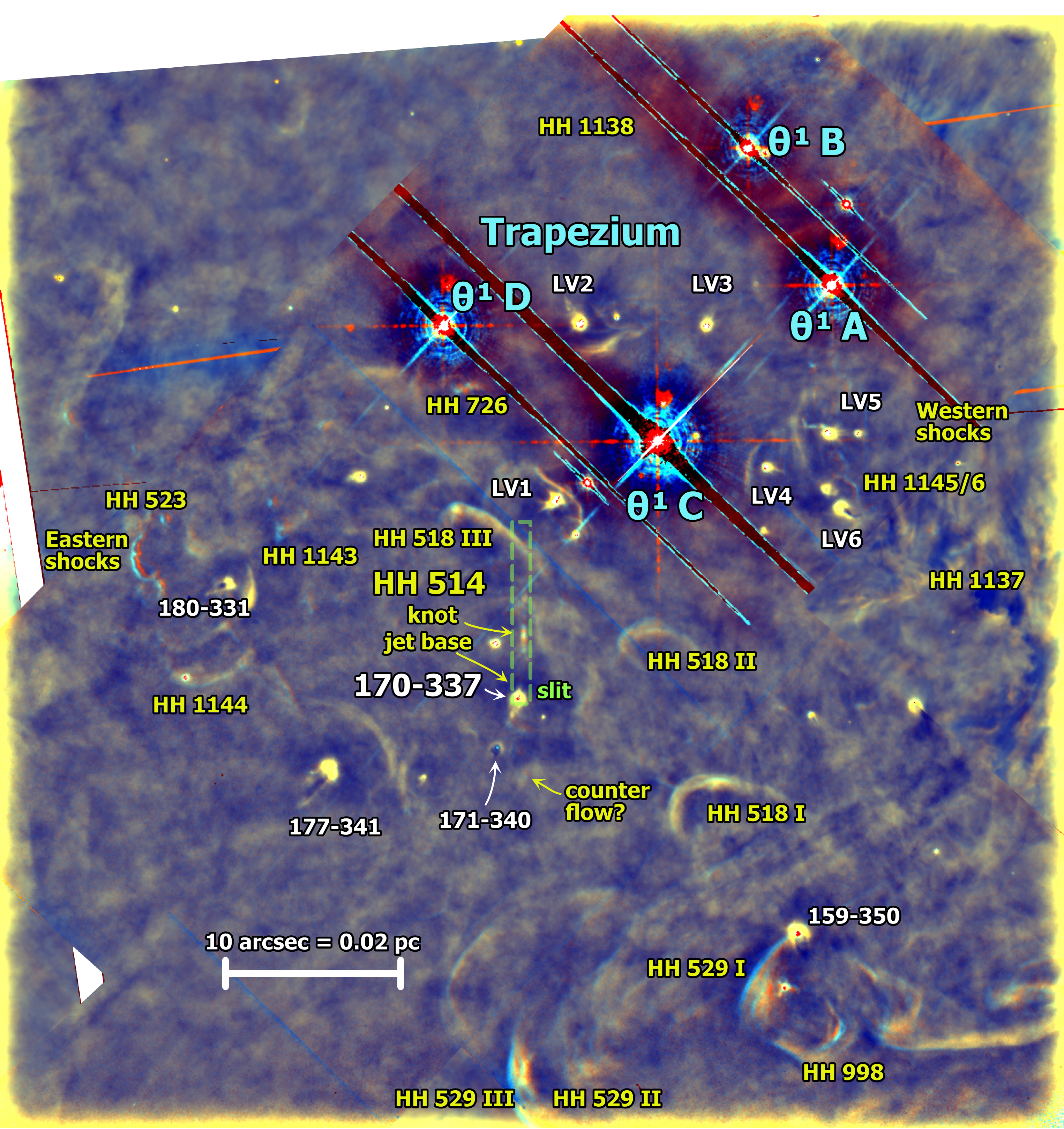}
\caption{
  Location of HH~514 within the inner Orion Nebula,
  showing the position and orientation of the UVES spectrograph slit
  (dashed green outline).
  High mass OB stars are labeled in blue,
  prominent proplyds are labeled in white,
  and components of Herbig-Haro flows
  from \citet{Odell15}
  are labeled in yellow.
  Background image is a composite of narrow-band observations
  with \textit{HST} in the [\ion{O}{3}] \(\lambda 5007\) line
  obtained at 3 different epochs over a span of 22 years:
  1994 (red channel),
  2004 (green channel),
  2016 (blue channel).
  All images are aligned to the astrometric frame defined in
  \citet{Robberto:2013a}
  and have been high-pass filtered with a Gaussian kernel of
  width 1~arcsec to suppress large-scale brightness variations.
  Stationary emission features appear yellow-white in this image,
  while moving features show a prism-like separation:
  red-green-blue, with blue indicating the leading edge.
}
\label{fig:hst}
\end{figure*}

Advances in astrophysical instrumentation in recent decades have allowed detailed observations of internal structures in Galactic star-forming regions. Among them, the observations of small-spatial scale structures in the Orion Nebula have been pioneering. Since the 1990s, with the first images from the Hubble Space Telescope (HST),
dozens of photoevaporating protoplanetary disks \citep[proplyds,][]{Odell1993}
have been discovered and analyzed in the nebula.
Proplyds generally have a compact semi-circular head
with an extended tail that points away from the ionizing source.
Their greatest concentration is in the central core of the Orion Nebula Cluster,
which contains the four high-mass Trapezium stars,
with \th1C{} (O7~Vp; \citealp{Sota:2011a}) being the most luminous member
and the dominant ionizing source for the nebula.
The Trapezium region is illustrated in Figure~\ref{fig:hst}
with an image that has been high-pass filtered to emphasize sub-arcsecond structures
(\(\SI{1}{arcsec} \approx \SI{0.002}{pc}\) for an assumed distance of \SI{410 \pm 10}{pc},
\citealp{Binder2018}).
High-mass Trapezium stars are labelled in blue
and proplyds are labelled in white following the nomenclature
of \citet{Laques:1979a} for the 6 closest proplyds to \th1C{} (LV knots)
and the coordinate-based names of \citet{Odell:1994a} for the remaining.

In many of the proplyds in the surrounding areas of \th1C,
supersonic gas outflows and jets have been observed.
Several of them have been classified as Herbig-Haro objects (HHs),
some of which are illustrated in Figure~\ref{fig:hst}.
This is the case of HH~514, which originates in the proplyd 170-337 \citep[][]{Bally:1998a},
located \SI{12.61}{arcsec} from \th1C.
A redshifted microjet of size \SI{0.5}{arcsec} protrudes northwards from the proplyd ionization front
and a short chain of redshifted knots
are seen at a distance of \(\approx \SI{4}{arcsec}\)
along the same axis \citep{bally00}
with proper motions consistent with an origin in 170-337 \citep{odellyhenney08}.
These are respectively labelled as ``jet base'' and ``knot'' in Fig.~\ref{fig:hst}.
A blueshifted counterflow is also detected at the base of the jet,
but it is much weaker.
This is unusual for HH flows in Orion,
which tend to be predominantly blueshifted \citep[][]{Henney:2007b}.
A faint fast-moving knot to the S of the proplyd may be associated with this counterjet.

HH~514 is particularly bright in the emission of [Fe\thinspace III] lines,
showing also intense lines of  [Fe\thinspace II], [Ni\thinspace II] and [Ni\thinspace III].
In addition, its proximity to $\theta^1$ Ori C and its large propagation velocity
make it an excellent laboratory to explore the phenomenon of destruction of dust grains
by analyzing the abundances of Fe and Ni in the gas.
Many HH objects have relatively large gaseous abundances of Fe and Ni,
whereas in the rest of the Orion Nebula these abundances are much smaller \citep[][]{mendez2021-2}.

In this paper, the third in a series dedicated to photoionized HH objects in the Orion Nebula,
we analyze for the first time the physical conditions, chemical composition and kinematics of HH~514.
We use high-resolution spectroscopy from the Ultraviolet and Visual Echelle Spectrograph (UVES) \citep[][]{Dodorico00}
of the Very Large Telescope
and IFU spectra from MEGARA \citep{gildepaz+18} at the Gran Telescopio Canarias (GTC).
Given the high spectral and spatial resolution of our observations,
we have analyzed two velocity components associated with two areas of HH~514:
the jet base and a northern knot, completely separated from the emission of the Orion Nebula.
Our analysis focuses mainly on estimating the total abundances of Fe and Ni,
which seems to be mainly in gaseous phase due to the strong degree of destruction of dust grains in the shock front.
We also report an overabundance of S and discuss its possible origin.

In Sec.~\ref{sec:data} we describe the observations and the data reduction process.
In Sec.~\ref{sec:physical_cond} we derive the physical conditions of each kinematic component.
In Sec.~\ref{sec:ionic_total_abundances} we derive the ionic and total abundances.
In Sec.~\ref{sec:disc} we discuss our results, focusing on the observed overabundance of S,
the total abundances of Fe and Ni and the kinematic properties of HH~514.
In Sec.~\ref{sec:summary_and_conclusions} we summarize our main conclusions.
Finally in the Appendix~\ref{sec:apendix_A} we add tables of data and figures as supporting material.

\section{Observations and data reduction}
\label{sec:data}

\subsection{High spectral resolution observations}
\label{sec:data_uves}

\begin{figure*}
  \begin{minipage}{6cm}
    \centering\includegraphics[height=2cm,width=\columnwidth]{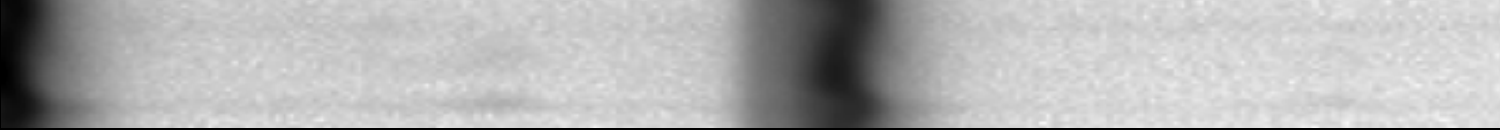}
    \centerline{(a) [O\thinspace II] $\lambda 3729$.}
    \smallskip
  \end{minipage}
  \begin{minipage}{6cm}
    \centering\includegraphics[height=2cm,width=\columnwidth]{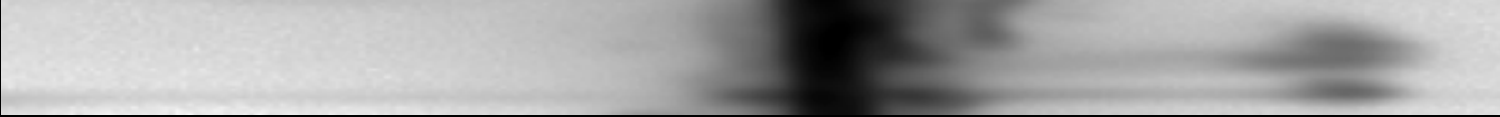}
    \centerline{(b) [O\thinspace III] $\lambda 4959$.}
    \smallskip
  \end{minipage}
 
  \begin{minipage}{6cm}
    \centering\includegraphics[height=2cm , width=\columnwidth]{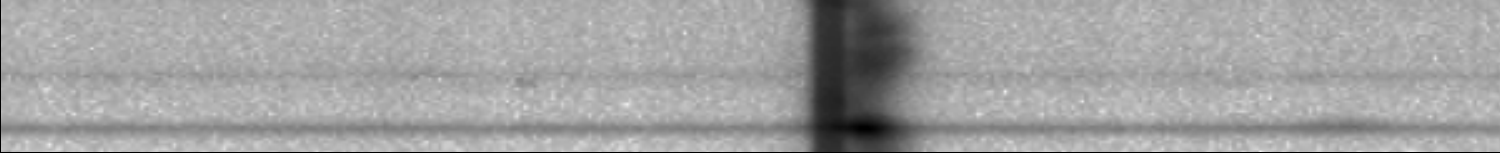}
    \centerline{(c) [O\thinspace I] $\lambda 6300$.}
    \smallskip
  \end{minipage}
  \begin{minipage}{6cm}
    \centering\includegraphics[height=2cm , width=\columnwidth]{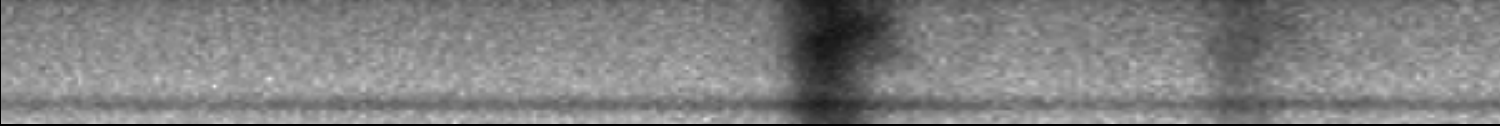}
    \centerline{(d) O\thinspace II $\lambda 4649$.}
    \smallskip
  \end{minipage}
\begin{minipage}{12.0cm}
  \centering\includegraphics[height=4cm, width=\columnwidth]{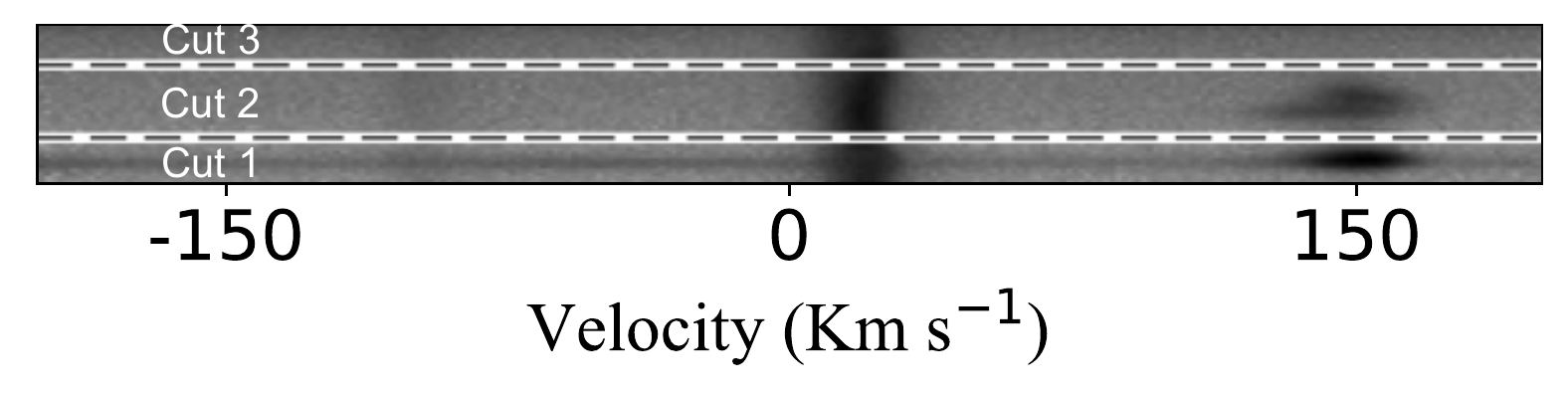}
  \centerline{(e) [Fe\thinspace III] $\lambda 4658$.} 
\end{minipage}

\caption{\textit{Upper and middle panels:} Sample of representative lines in the bi-dimensional spectrum. The Y axis corresponds to the spatial direction (up north, down south, see Fig.~\ref{fig:hst} for the spatial location of the slit) while the X axis is the spectral direction. All figures are centred at the rest-frame reference wavelength of each line. \textit{Bottom panel:} Emission of the [Fe\thinspace III] $\lambda 4658.17$ line as well as the limits and extension of the different spatial cuts extracted to analyse each velocity component. Cut 1 is at the bottom, which corresponds to the southernmost one. The spatial extension is 2.71 arcsec, 4.18 arcsec and 2.46 arcsec for cuts 1, 2 and 3, respectively. The velocity scale is heliocentric.}
\label{fig:cuts}
\end{figure*}

The observations were taken during the nights of 29 and 30 October, 2013, under clear conditions using the UT2 (Kueyen) of the Very Large Telescope (VLT) at Cerro Paranal, Chile, with the Ultraviolet Echelle Spectrograph \citep[UVES;][]{Dodorico00}. The central coordinates of the 10 arcsec slit are: RA(J2000)=05$^h$35$^m$16$^s$.95, DEC(J2000)=$-$05$^{\circ}$23$'$33.72$''$ oriented along the north-south spatial axis as is shown in Fig.~\ref{fig:hst}. The slit width of 1\arcsec provides an effective spectral resolution of $\lambda/\Delta \lambda \approx 6.5 \text{ km s}^{-1}$ in the spectral range between 3100 and 10400 \AA. Analogously to the observations described in the articles on HH~529~II-III and HH~204 \citep[][hereinafter Paper~I and Paper~II, respectively]{mendez2021,mendez2021-2}, three exposures of 150s each of the star GD71 \citep{Moehler14a, Moehler14b} were taken during the same night under similar conditions to achieve the flux calibration of the science data. The configuration of the instrument and the data reduction procedure are described in detail in Paper~I while the main settings of the observations of HH~514 are shown in Table~\ref{tab:obs_set}. We define three spatial cuts along the observed slit as shown in Fig.~\ref{fig:cuts}, covering the main high-velocity components,
which are redshifted by about $150\text{ km s}^{-1}$ from the nebular component.
We name the southernmost high-velocity component (within Cut~1) as the jet and the one located in the central cut as the knot. The nebular emission of Cut~1 also contains the emission of the proplyd 170-337. The complete analysis of the physical conditions and chemical composition of this proplyd will be the subject of a future work (M\'endez-Delgado et al., in prep). In this work we will only mention some results on the abundance of S in 170-337 in Sec.~\ref{subsec:real_overabundance}, which is relevant for our analysis of HH~514. Panel (b) of Fig.~\ref{fig:cuts} shows several components at intermediate redshifted velocities between that of the nebula and HH~514, which are seen mainly in lines of highly ionized ions such as [O\thinspace III] or [Ne\thinspace III]. Unfortunately, their emission is very weak in most spectral lines, so the determination of physical conditions and chemical abundances for those velocity components is not reliable with the present data and they are not studied further here.
They may be associated with the HH~518 flow from the proplyd 159-350
(see Figure~\ref{fig:hst}).
In order to increase the contrast of the HH~514 emission, which is rather weak compared to the main nebular background emission, we subtract Cut~3, which contains only background nebular emission, from Cut~1 and Cut~2, re-scaling its emission by the number of pixels of each cut. In addition to increasing the contrast of the HH~514 emission, this subtraction procedure permits us to remove sky and ghost contamination from the spectra of the HH object. We estimate the flux of the emission lines, their FWHM and their central wavelength by gaussian fittings using the SPLOT task of IRAF\footnote{IRAF is distributed by National Optical Astronomy Observatory, which is operated by Association of Universities for Research in Astronomy, under cooperative agreement with the National Science Foundation} \citep{Tody93} as is described in detail in Paper~I. We achieve the reddening correction by using the same procedure described in Paper~I and Paper~II, using the extinction curve from \citet{Blagrave07}. The resulting values of $\text{c}(\text{H}\beta)$, are presented in Table~\ref{tab:c_extin}. In Table~\ref{tab:sample_spectra} we include the relevant information for some of the emission lines of the spectrum of the jet base, as an example of the tables of the complete spectra that are provided as supplementary material.

\begin{table}
\caption{Main parameters of UVES spectroscopic observations.}
\label{tab:obs_set}
\begin{tabular}{ccccc}
\hline
Date & $\Delta \lambda$& Exp. time  &Seeing &Airmass\\
 & (\AA) &  (s) & (arcsec)&\\
\hline
2013-10-30 & 3100-3885 & 5, 3$\times$180 &0.92&1.11\\
2013-10-30 & 3750-4995 & 5, 3$\times$600 & 0.87 & 1.15\\
2013-10-30 & 4785-6805 & 5, 3$\times$180 &0.92&1.11\\
2013-10-30 & 6700-10420 & 5, 3$\times$600 & 0.87 & 1.15\\
\hline
\end{tabular}
\end{table}

\subsection{IFU observations}
\label{sec:data_megara}

Optical integral-field spectrograph observations of HH~514 were carried out on the night of 28 February 2022 under dark, photometric sky conditions and a seeing of $\sim$1.2 arcsec using MEGARA \citep{gildepaz+18} at the 10.4 m Gran Telescopio Canarias (GTC). The low-resolution (LR) Volume-Phased Holographic (VPH) grism VPH675-LR was used; this grism covers the 6095--7300 \AA\ wavelength range  with a spectral dispersion of 0.287 \AA\ pix$^{-1}$ and an effective resolution of R$\sim$5900. Three 900 s exposures were taken in order to minimize the impact of cosmic-rays in the relatively large field of view of MEGARA (12$''$.5 $\times$ 11$''$.3). 
The MEGARA data were reduced using the {\it megaradrp} v0.11 pipeline \citep{pascual+19, pascual+20}. The pipeline uses several python-based recipes to perform bias substraction, fibre-tracing,  flat-field correction, wavelength calibration, and spectra extraction. Sky subtraction was avoided owing to the large size of the Orion nebula prevented to place any of the MEGARA sky bundles on a nebular emission-free zone of the sky. Flux calibration was also performed using the spectrophotometric standard star HR\,1544. The final product is a row-stacked spectrum that is converted into a datacube of 0$''$.2 square spaxel on spatial dimensions using the {\it megararss2cube} task of the {\it megara-tools} suite v0.1.1 \citep{gildepaz+20}.

\begin{table}
\caption{Reddening coefficients for each component.}
\label{tab:c_extin}
\begin{tabular}{lccccc}
\hline
 & \multicolumn{2}{c}{$\text{c}(\text{H}\beta)$} \\
  & Nebular & High velocity\\
\hline
Cut 1 & - & $0.86 \pm  0.05$  \\
Cut 2 & $0.85 \pm 0.03$ &$0.75 \pm 0.07$\\
Cut 3 & $0.85 \pm 0.02$&-\\
\hline
\end{tabular}
\end{table}

\section{Physical Conditions}
\label{sec:physical_cond}

We use the version 1.1.13 of PyNeb \citep{Luridiana15} and the atomic data set shown in Tables 9 and 10 from Paper~II to derive the physical conditions of the different gas  components analyzed in this work. The adopted electron density, $n_{\rm e}$, in the nebular components is a weighted average\footnote{The weights were defined as the inverse of the square of the error associated to each density diagnostic.} of the resulting values from the CEL diagnostics
[O\thinspace II] $\lambda3726/\lambda3729$, [S\thinspace II] $\lambda6731/\lambda6716$, [Cl\thinspace III] $\lambda5538/\lambda5518$, [Fe\thinspace II] $\lambda9268/\lambda9052$, [Fe\thinspace III] $\lambda4658/\lambda4702$ and [Ar\thinspace IV]  $\lambda4740/\lambda4711$.  In addition, we apply the maximum-likelihood procedure described in Sec.~4.2 of Paper~I to derive the physical conditions based on the [Fe\thinspace III] $\lambda \lambda $ 4658, 4702, 4734, 4881, 5011 and 5271 lines. As it was commented in Paper~I, with the adopted atomic data, [Fe\thinspace III] diagnostics are not adequately sensitive to density for values smaller than $10^{3} \text{ cm}^{-3}$, thus if a considerable part of the observed ionized gas has densities of the order or lower than that value (as it is expected in the nebular components), the resulting $n_{\rm e}$ may be biased to that of the densest zones along the line of sight. The value of $n_{\rm e}(\text{O\thinspace II})$ was derived with the available RLs of multiplet 1 and the results are highly consistent with the adopted average density.

In the plasma diagnostics of the jet base, there are two possible intersections of the temperature sensitive curve of [N\thinspace II] $\lambda5755/\lambda6584$ with the density sensitive ones, giving very different values of the electron temperature,  $T_{\rm e}$. One gives an extremely high density (larger than 10$^5$ cm$^{-3}$) (shown in Fig.~\ref{fig:plasma}) or an extremely high temperature (higher than 10$^{5}$ K at smaller densities than 10$^{4}$ cm$^{-3}$, beyond the scale of Fig.~\ref{fig:plasma}). The [Fe\thinspace III] $\lambda4658/\lambda4702$, [S\thinspace II] $\lambda$4069/$\lambda$4076 and [O\thinspace II] $\lambda$7319+20/$\lambda$7330+31 density diagnostics confirm the first case. In the two studied components of HH~514 we estimate the convergence of $n_{\rm e}-T_{\rm e}$ by using the [Fe\thinspace III] $\lambda \lambda$ 4658, 4702, 4734, 4881, 5011, 5271 CELs. The resulting density in the jet base of HH~514 is $10^{5.37} \text{ cm}^{-3}$. 

In the case of the northern knot of HH~514, the convergence shows a higher dispersion in $n_{\rm e}-T_{\rm e}$ which seems to be due to the mixture of two or more gas components with similar velocities and notably different physical and/or ionization conditions. As it is shown in Fig.~\ref{fig:spatial_dis}, the spatial distribution along the slit of the intensity of the  [Fe\thinspace III] $\lambda 4658$ line centred at a heliocentric velocity of $\sim 150 \text{ km s}^{-1}$ is double-peaked in the northern knot of HH~514, which shows that at least two different gas components moving at a similar velocity are integrated in this part of the HH object. However, the separation into two spatial components results in a very low signal-to-noise spectra. Thus, the physical conditions derived in the knot are expected to be more uncertain than those in the jet base.

Based on the adopted density in each component, we estimate $T_{\rm e}$ through various diagnostics shown in Table~\ref{tab:pc}. In the nebular components, we define $T_{\rm e} \text{ (low)}$ as the weighted average of $T_{\rm e} (\text{[N\thinspace II]})$, $T_{\rm e} (\text{[O\thinspace II]})$ and $T_{\rm e} (\text{[S\thinspace II]})$ while $T_{\rm e} \text{ (high)}$ is the weighted average of $T_{\rm e} (\text{[O\thinspace III]})$, $T_{\rm e} (\text{[S\thinspace III]})$ and $T_{\rm e} (\text{[Ar\thinspace III]})$. In both components of HH~514, the estimation of $T_{\rm e} \text{ (low)}$ is quite uncertain, given the low critical density of the diagnoses available for this ionization zone and the high density of these objects. Due to this, and considering the similarity between $T_{\rm e}(\text{[O\thinspace III]})$ in both velocity components, we will adopt the $T_{\rm e} \text{ (low)}$ determined for the jet base --the spectrum with the best signal to noise ratio-- as also representative for the knot.


\begin{table*}
\centering
\caption{ Electron densities and temperatures determined from  several diagnostics.}
\label{tab:pc}
\begin{tabular}{ccccc}
\hline 
 & \multicolumn{1}{c}{Cut 1} & \multicolumn{2}{c}{Cut 2} & \multicolumn{1}{c}{Cut 3} \\
Diagnostic & HH~514~jet base & Nebula & HH~514~knot  & Nebula\\
\hline
& \multicolumn{4}{c}{$n_e$(cm$^{\text{-}3}$)}\\

[O\thinspace II] $\lambda$3726/$\lambda$3729 &  - &$5980^{+1160} _{-870}$&  - & $5610^{+960} _{-800}$\\

[O\thinspace II] $\lambda$7319+20/$\lambda$7330+31 &  263000: & - & - &- \\

[S\thinspace II] $\lambda$6731/$\lambda$6716 & - & $4150^{+1810} _{-1230}$&  -& $4240^{+1400} _{-1030}$\\

[S\thinspace II] $\lambda$4069/$\lambda$4076 & 219000: & - &-&-\\

[Cl\thinspace III] $\lambda$5538/$\lambda$5518 & - & $7640^{+970} _{-1020}$&  -& $7620^{+1070} _{-960}$\\

[Fe\thinspace II] $\lambda$9268/$\lambda$9052 & >76000 & $4500^{+9740} _{-3590}$ &  -& $3540^{+9190} _{-2760}$ \\

[Fe\thinspace III] $\lambda$4658/$\lambda$4702 & 234000: & $8490^{+2490} _{-2350}$& 57000: & $6430^{+2810} _{-2080}$\\

[Ar\thinspace IV]  $\lambda$4740/$\lambda$4711 & - &$6730^{+660} _{-690}$& - & $4800^{+570} _{-690}$\\

O\thinspace II$^{*}$  & - &$5640 \pm 510$&-&$4960 \pm 650$ \\

[Fe\thinspace III]$^{*}$ & $232000 \pm 12000$ & $8600 \pm 810$ & $74000 \pm 14000$& $7540\pm 1020$\\

\textbf{Adopted} &  \boldmath${232000 \pm 12000}$ &  \boldmath${6690 \pm 940}$&  \boldmath${74000 \pm 14000}$&  \boldmath${5490 \pm 1120 }$\\

 & \multicolumn{4}{c}{$T_e$ (K)}\\

T$\left(\mbox{He}\thinspace \mbox{I} \right)$ & $8710:$&$7790 ^{+570} _{-480}$ & $6010:$&$7570 ^{+490} _{-550}$\\

[N\thinspace II] $\lambda$5755/$\lambda$6584  & $12860^{+1240} _{-1170}$& $9750^{+230} _{-220}$&- &$9880^{+230} _{-250}$\\

[O\thinspace II] $\lambda \lambda$ 3726+29/$\lambda \lambda$7319+20+30+31 & $9180^{+1500} _{-630}$& $9500^{+740} _{-610}$& $7830^{+1430} _{-810}$&$10980^{+1040} _{-1320}$\\

[S\thinspace II] $\lambda \lambda$4069+76/$\lambda \lambda$ 6716+31& $20010^{+45030} _{-9150}$& $9410^{+950} _{-1220}$&-&$10090^{+2180} _{-1400}$\\

[O\thinspace III] $\lambda$4363/$\lambda \lambda$4959+5007& $8880^{+250} _{-210}$& $8490^{+90} _{-100}$ & $9330^{+260} _{-230}$& $8400^{+80} _{-70}$\\

[S\thinspace III] $\lambda$6312/$\lambda \lambda$9069+9531 & $8420^{+310} _{-400}$& $8630^{+330} _{-340}$& $8230^{+430} _{-560}$ & $8820^{+330} _{-310}$\\

[Ar\thinspace III]  $\lambda$5192/$\lambda$7136  & - & $8070^{+190} _{-160}$&-&$8320^{+230} _{-190}$\\

[Fe\thinspace III]$^{*}$ & $6840 \pm 1640$ & $8960 \pm 850$ & $8180 \pm 1410$& $7850 \pm 830$\\

\textbf{\boldmath${T_e}$ (low) Adopted} & \boldmath${11370 \pm 1820}$ & \boldmath${9530 \pm 100}$& \boldmath${11370 \pm 1820}$ & \boldmath${9920 \pm 200}$\\

\textbf{\boldmath${T_e}$ (high) Adopted} & \boldmath${8880 \pm 250}$ & \boldmath${8430 \pm 150}$ & \boldmath${9330 \pm 260}$ & \boldmath${8410 \pm 100}$\\

\hline
\end{tabular}
\begin{description}
\item $^*$ A maximum likelihood method was used. \\
\end{description}
\end{table*}



\section{Ionic and Total Abundances}
\label{sec:ionic_total_abundances}


We derive ionic abundances of N$^{+}$, O$^{+}$, S$^{+}$, Cl$^{+}$, Ca$^{+}$, Fe$^{+}$ and Ni$^{+}$ based on CELs and using $T_{\rm e} \text{ (low)}$. We derive the abundances of Cl$^{2+}$ and S$^{2+}$ using $T_{\rm e} \text{ ([S\thinspace III])}$ and the ionic abundances of O$^{2+}$, Ne$^{2+}$, Cl$^{3+}$, Ar$^{2+}$, Ar$^{3+}$ and Fe$^{3+}$ using $T_{\rm e} \text{ (high)}$. In the nebular components, we derive the abundance of Fe$^{2+}$ and Ni$^{2+}$ using $T_{\rm e} \text{ (low)}$. However, in both components of HH~514, the derived $T_{\rm e} \text{ ([Fe\thinspace III])}$ indicate a lower temperature than $T_{\rm e} \text{ (low)}$. This can be related to the dust destruction process described in detail in Sec.~\ref{subsec:real_overabundance}, where the filtering and destruction of dust grains increase the relative abundance of metals, cooling the gas. Therefore, in HH~514 we derive the abundances of Fe$^{2+}$ and Ni$^{2+}$ using $T_{\rm e} \text{ (high)}$. In the case of the abundance of Ca$^+$, we must take into account the possible coexistence of Ca$^+$ and H$^0$, due to its low ionization potential. Therefore, the derived Ca$^+$/H$^+$ may be an upper limit to the real abundance.

As discussed in Paper~II, some [Fe\thinspace II] and [Ni\thinspace II] lines can be produced mostly by continuum pumping instead of collisional excitation. We make sure that the estimation of the Fe$^+$ abundance is not affected by fluorescence by using lines of the lower  a$^4$F--a$^4$P levels  ($\lambda \lambda$ 7155, 8892, 9052, 9267) mostly populated by collisions \citep{Baldwin96}. In the case of Ni$^+$ the fluorescence effects may be larger, considering that the  levels involved in the transitions that give rise to the observed optical lines have the same multiplicity. However, if the density is high enough, collisional excitations may dominate the production of [Ni\thinspace II] lines. Following the same procedure described in equation 1 from Paper~II --based on the equation 8 from \citet{Bautista96}--, we estimate a critical density of $n_{\text{cf}}= 55860 \text{ cm}^{-3}$, from which collisions dominate over  fluorescence in the production of [Ni\thinspace II] $\lambda 7378$ line at the apparent distance of 12.61 arcsecs from $\theta^1$ Ori C. This implies that in both the jet base and the northern knot --regions with densities larger than $n_{\text{cf}}$--, we can directly calculate the abundance of Ni$^{+}$ based on the intensity of [Ni\thinspace II] $\lambda 7378$ line. However, this is not possible in the nebular components, where --due to the much lower density-- fluorescence should dominate.

In the UVES spectrum of the jet base of HH~514, we are able to estimate an upper limit for the abundance of Fe$^{3+}$ based on the noise level of the continuum at $\sim 6743.30$\AA, where we would expect the emission of [Fe\thinspace IV] $\lambda6740$. We carefully estimated this upper limit, taking into account that at 6742.91\AA~ there may be a relatively weak telluric emission line \citep[][]{Hanuschik03}. However, since Cut~3 was subtracted from Cut~1, --rescaling to the same number of pixels-- to increase the contrast between the nebular emission and the HH~514~jet base spectrum, any contamination by telluric lines has been eliminated. In this case we obtain an upper limit of 12+log(Fe$^{3+}$/H$^+$) $<$ 6.99.

\begin{figure}
\includegraphics[width=\columnwidth]{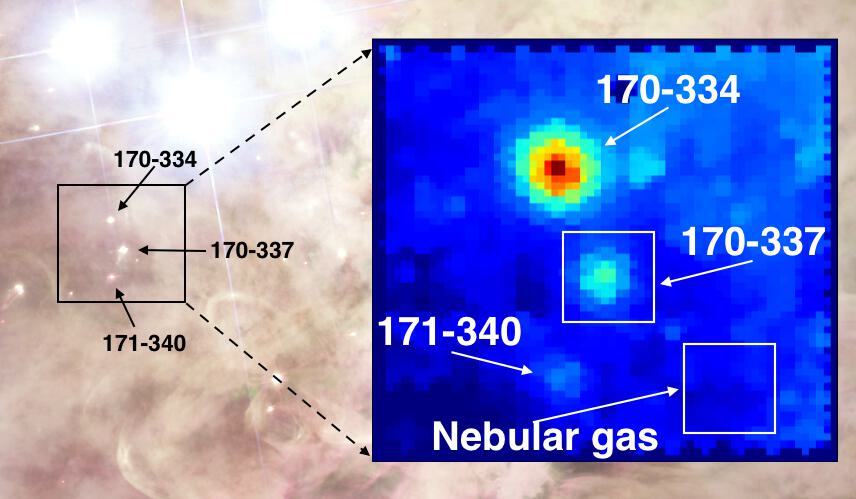}
\caption{ Location of the pointing of MEGARA-IFU observations. The field of view of MEGARA is 12$''$.5 $\times$ 11$''$.3 and the observations were centered in proplyd 170-337, where the high-velocity component corresponding to the jet base arise. White rectangles comprise the extracted spaxels to analyse the emission from proplyd 170-337 and those used for the adopted nebular background. The background HST image of the Orion Nebula is a RGB composite  \citep[Red: ACS F658N+F775W+F850LP; Green: ACS F555W; Blue: ACS F435W,][]{Robberto:2013a}.}
\label{fig:megara}
\end{figure}

\begin{figure}
\includegraphics[width=\columnwidth]{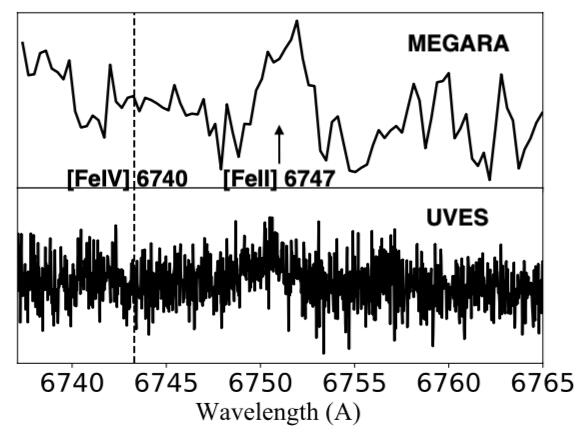}
\caption{Comparison between MEGARA and UVES one-dimensional spectra of the jet base of HH~514 between 6737-6765\AA. The flux units are normalized. In both cases the nebular background was subtracted. The expected location of the [Fe\thinspace IV] $\lambda 6740$ emission is indicated with a dashed vertical line. The depth of the MEGARA spectrum allows us to observe [Fe\thinspace II] $\lambda 6747$ as well as estimate a tighter upper limit to the [Fe\thinspace IV] $\lambda 6740$  emission.  }
\label{fig:megara_vs_UVES}
\end{figure}

To better constraint the flux of the [Fe\thinspace IV] $\lambda6740$ line in the jet base component of HH~514, we search into the MEGARA deep spectra cube (see Sect.~\ref{sec:data_megara}), analyzing the spatial area that comprises the proplyd 170-337 (which includes the emission of the jet base), subtracting the nebular emission from a surrounding area to optimize the contrast of the high-velocity emission -- as is indicated in Fig.~\ref{fig:megara} -- as well as remove weak sky features. The resulting 1D-spectrum is shown in Fig.~\ref{fig:megara_vs_UVES}, having a higher SNR than the UVES spectra, although with lower spectral resolution. This spectrum allowed us to restrict the upper limit the Fe$^{3+}$ abundance to   12+log(Fe$^{3+}$/H$^+$) $<$ 6.79.

In the nebular spectra, we estimate the He$^{+}$, C$^{2+}$, O$^{+}$, O$^{2+}$ and Ne$^{2+}$ abundances making use of RLs following the same procedure described in Paper~I. In the case of the two components of HH~514 analysed in this work, we can  only determine the abundances of He$^{+}$ from RLs. In this case, the adopted He$^{+}$ abundance is an average of the resulting values based on the available He\thinspace I singlet lines and triplet ones less affected by self-absorption effects (See Table D14 from Paper~I). The results are shown in Table~\ref{tab:ionic_abundances}. 

\begin{table*}
\centering
\caption{Ionic abundances derived in each kinematic component in logarithmic unit with $n(\text{H})=12$.}
\label{tab:ionic_abundances}
\begin{tabular}{ccccccccccccc}
\hline
 & \multicolumn{1}{c}{Cut 1} & \multicolumn{2}{c}{Cut 2} & \multicolumn{1}{c}{Cut 3} & \\
Ion &  HH~514~jet base & Nebula & HH~514~knot  & Nebula \\
\hline
 & \multicolumn{4}{c}{CELs}\\

N$^{+}$  & $6.88^{+0.28} _{-0.13}$ &$6.89 \pm 0.02 $  & $6.23^{+0.27} _{-0.14}$& $6.84 \pm 0.03 $\\

O$^{+}$ & $7.93^{+0.51} _{-0.18}$ & $7.81 \pm 0.03 $&$7.40^{+0.49} _{-0.19}$ & $7.69 \pm 0.05 $\\ 

O$^{2+}$ & $8.30^{+0.05} _{-0.04}$& $8.37 \pm 0.03$& $8.45^{+0.05} _{-0.04}$ &$8.38 \pm 0.02 $\\

Ne$^{2+}$ & $7.56^{+0.06} _{-0.05}$& $7.84^{+0.04} _{-0.03}$ & $7.79^{+0.06} _{-0.05}$&$7.82^{+0.03} _{-0.02}$\\

S$^{+}$ & $6.11^{+0.23} _{-0.12}$ & $5.50 \pm 0.04 $& $5.34^{+0.26} _{-0.14}$&$5.45 \pm 0.05 $\\

S$^{2+}$ & $7.39^{+0.07} _{-0.06}$ &$6.86^{+0.05} _{-0.04}$ & $7.35^{+0.09} _{-0.07}$&$6.84^{+0.05} _{-0.04}$\\

Cl$^{+}$ & - &$3.63 \pm 0.04 $&- &$3.60 \pm 0.04 $\\

Cl$^{2+}$ &- &$4.98^{+0.06} _{-0.05}$&- &$4.99^{+0.06} _{-0.05}$\\

Cl$^{3+}$ & - &$3.79 \pm 0.03 $&- &$3.92 \pm 0.03 $\\

Ar$^{2+}$ & $6.29 \pm 0.04 $& $6.31 \pm 0.03 $& $6.19 \pm 0.04 $&$6.31 \pm 0.02 $\\

Ar$^{3+}$ & - &$5.09^{+0.04} _{-0.03}$& -&$5.15^{+0.03} _{-0.02}$\\

Ca$^{+}$ & $3.71^{+0.20} _{-0.12}$ &- &-&-\\

Fe$^{+}$ & $6.14 \pm 0.05$ & $4.39 \pm 0.05$ & - &$4.32 \pm 0.02$\\ 

Fe$^{2+}$ & $7.14 \pm 0.02$ & $5.28 \pm 0.01$& $6.83 \pm 0.04$&$5.17 \pm 0.02$\\

Fe$^{3+}$ & < 6.79 &$5.62^{+0.10} _{-0.08}$& -&$5.78 \pm 0.12 $\\

Ni$^{+}$ & $4.94^{+0.15} _{-0.09}$ & - &-&-\\

Ni$^{2+}$ &$5.84^{+0.07} _{-0.06}$ & $4.32 \pm 0.03 $ &$5.62^{+0.08} _{-0.07}$&$4.25 \pm 0.04 $\\

 & \multicolumn{4}{c}{RLs}\\

He$^{+}$  &$10.95 \pm 0.04$  & $10.92 \pm 0.02$& $10.96 \pm 0.02$&$10.92 \pm 0.02$\\

C$^{2+}$  & -& $8.37 \pm 0.01 $ &-&  $8.37 \pm 0.01$ \\

O$^{+}$ & - & $8.17^{+0.04} _{-0.03}$& -& $8.19 \pm 0.05 $ \\

O$^{2+}$ & - & $8.61 \pm 0.05$ & - & $8.63 \pm 0.05$  \\

Ne$^{2+}$ & -&$8.04 \pm 0.06 $&-& $8.10 \pm 0.08 $ \\

\hline
\end{tabular}
\end{table*}


To estimate the total abundances of C, N, Ne, S and Ar, we use the ionization correction factors (ICFs) from \citet{Amayo2021}. These ICFs, based on photoionization models of giant H~II regions, are consistent with the schemes shown in table~10 of Paper~I. However, they additionally provide the 1-$\sigma$ limits of their predictions, allowing the formal propagation of uncertainties associated to the ICFs. In our nebular spectra, the calculation of the total abundances of O, Cl and Fe did not require ICF and were estimated from the sum of their ionic abundances. In the case of Ar, Ar$^+$ is estimated to contribute only 1 or 2 per cent of the total Ar abundance, having a negligible impact. We estimate the total abundance of Ni as $\text{Fe/Ni}=\text{Fe}^{2+}/\text{Ni}^{2+}$ based on their similar depletion and ionization patterns \citep[][]{mendez2021-2}. In the case of the He abundance in the nebular components, we use the ICF of \citet{kunthsargent83}. Since we have a reliable determination of the abundance of O$^+$ and O$^{2+}$ both with CELs and RLs, we derive the total abundances of He, Ne and C using the degree of ionization estimated from the abundances of the two ionic species of O determined from RLs.

In the case of the jet base of HH~514, we determine the total abundance of Fe as Fe/H = Fe$^{+}$/H$^{+}$+Fe$^{2+}$/H$^{+}$+Fe$^{3+}$/H$^{+}$, summing the estimated upper limit of 12+log(Fe$^{3+}$/H$^{+}$) = 6.79. At first approximation, the total abundance of Fe should be between $7.18 \pm 0.02$ and $7.33 \pm 0.02$, where the value of 12+log(Fe$^{3+}$/H$^{+}$) takes values between 0 and 6.79.  \citet{rodriguez05} provides two ICF(Fe) schemes, where the true Fe abundance should be between the limits predicted by both schemes, based on the observed O$^{+}$/O$^{2+}$ and Fe$^{2+}$/O$^{+}$ ratios. In our case, both ICFs match in the value of 7.50. Therefore, considering the observational uncertainties in the abundance of O$^{+}$, this suggest that the upper limit of the value of Fe$^{3+}$/H$^{+}$ should be close to its expected true value and therefore 12+log(Fe/H)$\approx 7.33 \pm 0.02$. 

It is remarkable that log(Ni$^{+}$/Fe$^{+}$) = $-1.20 \pm 0.16$, while log(Ni$^{2+}$/Fe$^{2+}$) = $-1.30 \pm 0.07$, which confirms the conclusions of \citet{mendez2021-2} about the similar ionization and depletion patterns of Ni and Fe. Furthermore, this demonstrates that the effects of fluorescence on our Ni$^{+}$ abundance determination are negligible, since we would expect stronger continuum pumping effects for [Ni\thinspace II] than [Fe\thinspace II]. The estimated total abundances are shown in Table~\ref{tab:total_abundances}.


\begin{table*}
\centering
\caption{Total abundances. The units are logarithmic with $n(\text{H})=12$.}
\label{tab:total_abundances}
\begin{tabular}{ccccccccccccc}
\hline
 & \multicolumn{1}{c}{Cut 1} & \multicolumn{2}{c}{Cut 2} & \multicolumn{1}{c}{Cut 3} & \\
Ion &  HH~514~jet base & Nebula & HH~514~knot  & Nebula \\
\hline
 & \multicolumn{4}{c}{CELs}\\

N  & $7.43 ^{+0.59} _{-0.32}$ & $7.64 \pm 0.16$ &$7.36 ^{+0.58} _{-0.31}$ & $7.69 ^{+0.18} _{-0.16}$ \\ 

O  & $8.45 \pm 0.16$ & $8.48 \pm 0.02$ &$8.49 \pm 0.06$&$8.46 \pm 0.02$\\

Ne & $7.75 ^{+0.19} _{-0.15}$ & $7.97 ^{+0.11} _{-0.10}$& $7.84 ^{+0.11} _{-0.09}$ & $7.92 ^{+0.09} _{-0.08}$ \\ 

S  & $7.43 ^{+0.16} _{-0.12}$ & $6.92 ^{+0.07} _{-0.06}$& $7.49 ^{+0.17} _{-0.14}$ &$6.91 ^{+0.07} _{-0.06}$\\

Cl & - & $5.03 \pm 0.05$&- &$5.04 \pm 0.04$\\

Ar & $6.32 ^{+0.12} _{-0.10}$ & $6.34 ^{+0.06} _{-0.05}$ & $6.29 ^{+0.14} _{-0.12}$& $6.34 \pm 0.04$\\

Fe & $7.33 \pm 0.02$ & $5.80 \pm 0.07$ & - &$5.89 \pm 0.09$\\


Ni & $6.03 \pm 0.08$ & $4.83 \pm 0.08 $&-&$4.96 \pm 0.10$\\

 & \multicolumn{4}{c}{RLs}\\

He  &$10.95 \pm 0.04$  & $10.95 \pm 0.02$ & $10.96 \pm 0.02$ & $10.95 \pm 0.02$\\

C & - &$8.45 ^{+0.07} _{-0.06}$&-&$8.45 ^{+0.07} _{-0.06}$\\ 

O  &  - & $8.74 \pm 0.04$ &-&$8.76 \pm 0.04$\\ 

Ne  & - &$8.17 ^{+0.13} _{-0.11}$&-&$8.26 ^{+0.18} _{-0.14}$\\

\hline
\end{tabular}
\end{table*}

\section{Discussion}
\label{sec:disc}

\subsection{The star/disk/outflow system of 170-337/HH~514}
\label{sec:general-properties}

In this section we combine results from the literature with those of this study
to provide a comprehensive view of the multiple components
that make up the proplyd 170-337 and its outflows. 

\subsubsection{The central star and its disk}
\label{sec:central-star-disk}

The central star of 170-337 is of spectral type M2e \citep{Hillenbrand:1997a}
and its mass and age have been determined from fitting to
pre-main-sequence evolutionary tracks,
yielding \SI{0.3}{\msun} and \SI{0.6}{Myr} respectively \citep{Boyden:2020a}.
It was one of the first Orion proplyds to be detected
via the mm-wave dust emission of its protostellar disk \citep{Williams:2005a},
with an estimated dust mass of \SI{23 \pm 5}{\mearth} \citep{Eisner:2018a}.
The disk has also been detected in the CO 3--2 line \citep{Boyden:2020a},
yielding a disk radius of \SI{44 \pm 4}{AU},
which is roughly half of the radius of the proplyd ionization front
(\SI{90}{AU}, \citealp{Henney:1998b}).
Unfortunately, the gas mass cannot be determined from the CO observations
because the line is optically thick.

\subsubsection{Launching regions of the jet and the photoevaporation flow}
\label{sec:launch-regi-jet}

No binary companion to the central star of 170-337 has been detected,
either spectroscopically \citep{Tobin:2009a}
or visually \citep{Duchene:2018a}.
Therefore, it is likely that
both the proplyd and HH flows originate from the same protostellar disk,
but the flows must be launched from very different radii.

In the case of the proplyd, an initially neutral photoevaporation flow
is driven from the disk surface by FUV radiation from the Trapezium stars
\citep{Johnstone:1998a}.
For the high radiation fields found in the Orion Nebula,
the flow is in the supercritical regime \citep{Adams:2004u}
in which the sound speed in the FUV-heated gas (\(\approx \SI{3}{km.s^{-1}}\))
exceeds the local gravitational escape velocity at the disk surface.
For a stellar mass of \SI{0.3}{\msun} (sec.~\ref{sec:general-properties}),
this implies radii greater than about \SI{30}{AU}
and up to the outer radius of the disk (\SI{44}{AU}).

In the case of the jet,
the much higher flow velocity (\(> \SI{100}{km.s^{-1}}\))
implies a launch region that is deeper in the gravitational well
and closer to the star.
Comparison of magneto-centrifugal jet models
with observations of the base of HH~212 \citep{Lee:2017r}
imply that jet launching occurs at disk radii \(< \SI{0.1}{AU}\).
Magnetocentrifugal winds may also be launched from intermediate radii
of \num{1} to \SI{10}{AU} but these will be at lower velocities
(e.g., \citealp{Hasegawa:2022j}).

\subsubsection{Kinematics of the star within the Orion Nebula Cluster}
\label{sec:kinem-star-with}

The line-of-sight radial velocity of the star can be determined from the systemic velocity of the CO emission
\citep{Boyden:2020a},
yielding \SI{23.8 \pm 0.1}{km.s^{-1}} after converting to the heliocentric frame.
This is  \SI{-3.7}{km.s^{-1}} with respect to the mean velocity of stars in the core of Orion
(\SI{27.5}{km.s^{-1}}, \citealp{Theissen:2022k}).
The star's tangential velocity in the frame of the cluster can be derived from proper motion measurements \citep{Kim:2019a}
as \(\SI{7.8 \pm 1.1}{km.s^{-1}}\) along \(\text{PA} = \ang{48 \pm 8}\).
The total space velocity of 170-337 is therefore
\SI{8.6}{km.s^{-1}},
which is more than double the 3D velocity dispersion of the cluster
(\SI{3.7}{km.s^{-1}}; \citealp{Theissen:2022k}),
making it a candidate for an escaping star.
It is included in the escape group ESC1 of \citet{Kim:2019a},
but its total proper motion of \SI{4}{mas.yr^{-1}}
is not quite high enough to meet the runaway criterion
used in \citet{McBride:2019a} or \citet{Platais:2020f}.

\subsubsection{Kinematics of the jet}
\label{sec:kinematics-jet}

Proper motion measurements of the jet knot \citep{odellyhenney08},
after compensating for the proper motion of the star and
rescaling to our adopted distance of \SI{410}{pc},
yield a plane-of-sky tangential velocity of \SI{41 \pm 6}{km.s^{-1}},
with a position angle that is consistent with the displacement
of the knot from 177-341 (almost due north).
Line-of-sight radial velocities can be determined from our spectroscopic measurements
(see Table~\ref{tab:sample_spectra})
and are found to be approximately the same for the jet base and the knot,
with no variation between high and low ionization lines,
yielding a heliocentric velocity of \SI{+150 \pm 5}{km.s^{-1}}.
Subtracting the stellar systemic velocity of \SI{24}{km.s^{-1}}
gives a jet propagation velocity of  \SI{132 \pm 10}{km.s^{-1}}
at an inclination of \ang{-72 \pm 3} to the plane of the sky.
Note that the brighter northern side of the flow is inclined away from the observer.

Assuming an ionized sound speed of \SI{11}{km.s^{-1}},
then the isothermal Mach number of the jet is \num{12}
and the Mach angle is \ang{4.8},
which should correspond to the half-opening angle of the ionized portion of the jet
after it emerges from the proplyd ionization front.
This is consistent with the observed width of the base of the jet in \textit{HST} images
(e.g., Fig.~\ref{fig:hst}).

\subsubsection{Mass loss rates of photoevaporation flow and jet}
\label{sec:mass-loss-rate}

The mass-loss rate from the proplyd photoevaporation flow is estimated
from model fits to kinematic line profiles \citep{henney99}
to be \(\approx \SI{8e-7}{\msun.yr^{-1}} \),
with an uncertainty of roughly a factor of two.
\citet{Bally:2006a} include HH~514 in a survey of irradiated jets in the Orion Nebula,
calculating a jet mass loss rate of \SI{6e-7}{\msun.yr^{-1}},
which would make it one of the most massive jets in their survey.
However, this estimate is based on analysis of the jet knot and therefore
corresponds to a maximum value of the mass loss,
presumably during an outburst phase.
Assuming that the current conditions at the jet base are more typical,
then a time-averaged jet mass loss can be estimated by scaling the
result from the proplyd photoevaporation flow.
Since the densities at the base of the jet and in the proplyd flow are similar
(section~\ref{sec:physical_cond} and further details in \ref{subsec:proplyd} below),
the mass loss rate scales as the flow velocity times its solid angle.
Assuming a jet half opening angle of \ang{5} (see previous section)
implies a mean jet mass loss rate
of \num{0.05} times that of the proplyd,
yielding \SI{4e-8}{\msun.yr^{-1}} albeit with a large uncertainty.
This is similar to but slightly larger than the jet mass-loss rate
derived for the closer-in proplyd LV~2 (167-317, \citealp{Henney:2002b}).

\subsubsection{Discrepancy between optical jet orientation and CO kinematics}
\label{sec:discr-betw-optic}

Assuming that the jet is perpendicular to the disk,
then the results of sec.~\ref{sec:kinematics-jet} imply that the disk must be close to face-on.
In the disk literature, a different inclination convention is used
in which the \(i\disk\) is the angle between the rotation axis and the line of sight,
meaning that our jet results imply \(i\disk = \ang{18}\) for 170-337.
Studies of the mm continuum and line emission \citep{Eisner:2018a, Boyden:2020a}
find that the disk has a slight elongation in the north-south direction,
which is interpreted as \(i\disk \approx \ang{40}\),
but if the jet results are taken at face value
then this elongation must be an intrinsic asymmetry rather than an inclination effect. 
\citet{Boyden:2020a} derive a very low stellar mass (\(< \SI{0.1}{\msun}\))
from kinematic disk modelling.
However, this value is highly dependent on the disk inclination
(\(M \sim (\sin i\disk)^{-2}\)),
so that rescaling to \(i\disk = \ang{18}\) would give \SI{0.3}{\msun},
which is fully compatible with the mass derived from stellar evolution tracks
(see sec.~\ref{sec:central-star-disk} above).

A more serious conflict with the CO kinematics concerns the
sign and orientation of the observed velocity gradient,
which runs north-south with blueshifted CO on the north side
(Figs.~4 and~11 of \citealp{Boyden:2020a}).
This is inconsistent with an origin either in a rotating disk
or in an outflow aligned with the optical jet.
In the first case, the disk rotation axis would have to be perpendicular to the jet axis.
In the second case, one would expect \emph{red}shifted instead of blueshifted CO on the north side.
Two possible explanations for this discrepancy suggest themselves:
(1)~the source, despite lack of evidence, may in fact be binary,
with the optical jet arising from a different star/disk from that seen in CO;
(2)~the CO emission might arise in neither the disk nor the outflow,
but instead in in-falling material from the remnants of the surrounding protostellar envelope.
Both possibilities seem far-fetched,
but in the second case there is a precedent in observations of larger Orion proplyds,
where dense neutral filaments are seen within the shadow of the ionization front.
Examples include 182-413 \citep{Vicente:2013a} and 244-440 \citep{bally00},
which are likely to represent transitional stages between Evaporating Gaseous Globules
\citep{Hester:1996w}
and pure photoevaporating disks.
Although 170-337 is much smaller, it too may contain remnant envelope material
that is still accreting onto the central source.
The same phenomenon may explain asymmetrical dust extinction features
that are seen in the tails of many proplyds,
such as 171-340 (see Fig.~\ref{fig:hst}).

\subsection{On the anomalous [S\thinspace III] emission in HH~514}
\label{subsec:under_TS3}

There are two results for HH~514 included in Tables \ref{tab:pc} and \ref{tab:total_abundances} that we want to highlight: both components show $T_{\rm e} \text{ ([S\thinspace III])}$ $<$ $T_{\rm e} \text{ ([O\thinspace III])}$, and their S/H ratio is around 0.5 dex higher than the value found in the nebular gas. The abnormally large S abundance seems to rely mainly on the S$^{2+}$ one, which represents 91 per cent of the S/H ratio.

The observed trend of $T_{\rm e} \text{ ([S\thinspace III])}$ goes in the opposite direction to the usual temperature stratification observed in H~II regions and simple photoionization models fail to reproduce the observed results \citep[][]{Binette2012}. On the other hand, the overabundance of S is much too large to be produced by errors in the assumed ICF (the abundance of S$^{2+}$ alone is higher than the total S abundance of the Orion Nebula).

In order to be certain that the enhanced S abundance is real,
it is necessary to rule out possible systematic errors in the analysis.
To that end, in the following subsections we explore some scenarios that could explain the observed results.

\subsubsection{Errors in atomic data for analyzing [S\thinspace III] lines?}
\label{subsec:atomic_data}

After analyzing different sets of transition probabilities and collision strengths of different ions, \citet{Juan-de-Dios17} conclude that the derived ionic abundances can have uncertainties of around 0.1-0.2 dex introduced by the atomic data when density is lower than $10^4$ cm$^{-3}$, but they can reach values up to 0.8 dex for higher densities. Since both components of HH~514 have extremely high densities, we expect the uncertainties due to atomic data may be amplified in these objects. Given such expectations, we decided to explore the effect of the atomic data used on the S$^{2+}$ abundance.

Instead of following the methodology of  \citet{Juan-de-Dios17} of analyzing all possible combinations of transition probabilities and collision strengths available, we first focus our attention on transition probabilities, discarding those that give clearly wrong results. If we do not do this in the first place, we may include spurious values in the resulting abundance distribution, which would increase the uncertainties associated with the atomic data. Once this first step is done, we  check the available collision strengths, analyzing their impact on $T_{\rm e} \text{ ([S\thinspace III])}$ and the S$^{2+}$ abundance.


We use the observed intensity of [S\thinspace III] $\lambda \lambda 9531, 9069, 8829$ lines, arising from the common $^1$D$_2$ upper level and the [S\thinspace III] $\lambda \lambda 3722, 6312$ lines, emitted from the same $^1$S$_0$ level. Their intensity ratios depend exclusively on the transition probabilities. In Table~\ref{tab:atomic_data_test}, we compare the aforementioned [S\thinspace III] line intensity ratios observed in several positions and HH objects of the Orion Nebula and other Galactic H\thinspace II regions for which we have deep high-spectral resolution spectra with the line ratios predicted by the different atomic datasets.    

As it is well known, infrared [S\thinspace III] $\lambda \lambda 9531, 9069$ lines can be affected by telluric absorption bands \citep{Noll12}. In \citet{mendez2021} and \citet{mendez2021-2}, we mentioned that [S\thinspace III] $\lambda 9069$ was affected by telluric absorptions, while $\lambda 9531$ remained unaffected in the background emission of the Orion Nebula, while in the HHs both lines avoided the absorption bands. Since in this work we are interested in analyzing the [S\thinspace III] $\lambda$9531/$\lambda$9069 line ratio, we correct the spectra from telluric absorptions around the aforementioned lines. We use the observations of the star GD71 from each data set, normalizing them with the standard tabulated flux. Then, the science spectra were divided by the resulting telluric transmission function around [S\thinspace III] $\lambda \lambda 9531, 9069$. The corrected intensities are shown in Table~\ref{tab:tellabs}. 

[S\thinspace III] $\lambda 3722$ is usually blended with the H14 Hydrogen Balmer line. We can not simply subtract the emission of H14 using its theoretical line ratio with respect to other isolated observed H\thinspace I lines, since it is well known that H\thinspace I lines from high quantum numbers $n$ suffer deviations from case ``B'' \citep[][]{mesadelgado09, rodriguez20}. Even using  contiguous lines such as H13 or H15 may introduce important deviations \citep[see fig.~A2 of][]{rodriguez20}. Therefore, we consider that it is better to use only the observations where the spectral resolution of the data allows the correct Gaussian deblending of the lines, where the resulting FWHM of [S\thinspace III] $\lambda 3722$ is consistent with that of the isolated [S\thinspace III] $\lambda 6312$ line.

The transition probabilities tested are those available in version 1.1.13 of PyNeb and those from the CHIANTI database V10 \citep[][]{Chianti10}, named by the following acronyms: LL93-HSC95-MZ82b-KS86 \citep[][]{LL93,HSC95,MZ82b, KS86}, MZ82b-HSC95-LL93\citep[][]{MZ82b,HSC95,LL93}, FFTI06 \citep[][]{FFTI06}, TZS19\citep[][]{TZS19} and CHIANTI \citep[][]{Tayal97,FFTI06, Hudson12}. These data encompass all those studied by \citet{Juan-de-Dios17} adding  those from TZS19 and CHIANTI. The FFTI06 data file  was named as PKW09 in previous versions of PyNeb, alluding to \citet{Podobedova09}, who cited \citet{FFTI06}. 

As shown in Table~\ref{tab:atomic_data_test}, the LL93-HSC95-MZ82b-KS86 set of  transition probabilities are clearly inconsistent with observations when determining line ratios involving the [S\thinspace III] $\lambda 9531$ line from the $^1$D$_2$-$^3$P$_2$ transition, for which atomic data are taken from \citet{KS86}. This discrepancy forces us to discard   the LL93-HSC95-MZ82b-KS86 data set. With respect to the other data sets, the ones most consistent  with the observed line ratios are those by FFTI06 and CHIANTI. However, as is shown in Fig.~\ref{fig:A_sii}, the results obtained with the remaining data sets are not very different for the most important optical transitions ([S\thinspace III] $\lambda \lambda 3722, 6312, 8829, 9069, 9531$). In fact, taking the mean and standard deviation of the A-values given by the datasets included in Fig.~\ref{fig:A_sii} excluding those from LL93-HSC95-MZ82b-KS86, we obtain A$_{3722}=(7.06 \pm 0.53) \times 10^{-01}$, A$_{6312}=2.20 \pm 0.08$, A$_{8829}=(6.42 \pm 1.49) \times 10^{-06}$, A$_{9069}=(2.00 \pm 0.13) \times 10^{-02}$ and A$_{9531}=(5.26 \pm 0.35) \times 10^{-02}$. Of these, the one with the largest deviation is A$_{8829}$, about 23 per cent. In any case, this transition probability has very little impact in the determination of $T_{\rm e} \text{ ([S\thinspace III])}$ and the S$^{2+}$ abundance if a line different than [S\thinspace III] $\lambda 8829$ is used, which is always the case. Moreover, if we discard the dataset of TZS19, this deviation is reduced to 4 per cent. Once the LL93-HSC95-MZ82b-KS86 data set is removed, the uncertainties associated with the transition probabilities of the optical lines seem to be around $\sim 5$ percent considering the consistency between the predictions and the observed line intensity ratios included in  Table~\ref{tab:atomic_data_test}. The previous results indicate that the available sets of transition probabilities for transitions giving rise to [S\thinspace III] lines show a high degree of consistency. From now on, we adopt the transition probabilities of FFTI06 in our calculations. This was precisely the data set we used in Paper~I and Paper~II.

Analogously to the previous procedure with transition probabilities, we also check the data sets of collision strengths available in PyNeb: GMZ95 \citep[][]{GMZ95}, TG99 \citep[][]{TG99}, HRS12 \citep[][]{HRS12}, GRHK14 \citep[][]{GRHK14} and TZS19 \citep[][]{TZS19}. These includes all the sets analyzed by \citet{Juan-de-Dios17} in addition to TZS19. The CHIANTI data base V10 adopts the same collision strengths as HRS12 by default, so does not need to be considered separately. Comparing the results for the different data sets shown in Fig.~\ref{fig:omega_siii_zoom }, we obtain the following mean collision strengths along  with their associated dispersion at 10,000 K for the most important [S\thinspace III] lines: $\Omega_{3722}=(3.65 \pm 0.31) \times 10^{-1}$, $\Omega_{6312}=1.38 \pm 0.04$, $\Omega_{8829}=(7.64 \pm 0.63)\times 10^{-1}$, $\Omega_{9069}=2.28 \pm 0.20$ and $\Omega_{9531}=4.00 \pm 0.26$. The typical dispersion is about $\sim 8$ per cent for most lines, except for [S\thinspace III] $\lambda 6312$, where the dispersion is smaller, $\sim 3$ per cent.

One way in which errors in the collision strengths could explain
the apparent low $T_{\rm e} \text{ ([S\thinspace III])}$ and the high S$^{2+}$ abundance in the HH objects
is by changing the  $\frac{\Omega_{\text{Auroral}}}{\Omega_{\text{Nebular}}}=\frac{\Omega(3722)+\Omega(6312)}{\Omega(8829)+\Omega(9069)+\Omega(9531)}$ ratio. When $\frac{\Omega_{\text{Auroral}}}{\Omega_{\text{Nebular}}}$ decreases, $T_{\rm e} \text{ ([S\thinspace III])}$ increases, for a fixed observed [S\thinspace III] $\frac{I(\lambda3722)+I(\lambda6312)}{I(\lambda 8829)+I(\lambda9069)+I(\lambda9531)}$ ratio. Increasing the value of the denominator in the ratio of collision strengths also decreases the derived abundance of S$^{2+}$, which is usually determined from the intensity of [S\thinspace III] $\lambda \lambda 9531, 9069$ lines. However, modifications in the values of $\Omega_{8829}, \Omega_{9069} \text{ and } \Omega_{9531}$ within a dispersion of $\sim8$ per cent are not enough to account for the discrepant values of $T_{\rm e} \text{ ([S\thinspace III])}$ and the S$^{2+}$ abundance.

We will try to solve the aforementioned discrepancies exploring the assumption that the collision strengths of the transitions that give rise to [S\thinspace III] lines are wrong, with $\Omega_{9531}, \Omega_{9069}$ and $\Omega_{8829}$ being underestimated beyond the dispersion of $\sim8$ per cent. We will increase the values of $\Omega_{9531}, \Omega_{9069}$ and $\Omega_{8829}$, until getting $T_{\rm e} \text{ ([O\thinspace III])}=T_{\rm e} \text{ ([S\thinspace III])} $ in the jet base of HH~514. This may be considered a lower limit since it is expected that $T_{\rm e} \text{ ([S\thinspace III])} $ is between $T_{\rm e} \text{ ([O\thinspace III])} $ and $T_{\rm e} \text{ ([N\thinspace II])} $ \citep[][]{Berg20}, the latter being generally higher. To achieve this, the collision strengths should be increased about 40 per cent, as we know, 5 times larger than the dispersion found comparing the available atomic data. In Table~\ref{tab:modified_data_omega} we show the $T_{\rm e} \text{ ([S\thinspace III])} $ and  S$^{2+}$ abundances we obtain for each object using the transition probabilities of FFTI06, the different sets of collision strengths considered and the corresponding value of $n_{\rm e}$. In the last column, we include the results obtained increasing by a 40 per cent the $\Omega_{9531}, \Omega_{9069}$ and $\Omega_{8829}$ values of GRHK14. 

Now we discuss what other consequences the increase of 40 percent in the collision strengths may have. Using these $\Omega$ modified values, $T_{\rm e} \text{ ([S\thinspace III])}$ matches $T_{\rm e} \text{ ([O\thinspace III])}$ in both components of HH~514\footnote{This is obviously expected for the jet base, since the increase in the collision strengths was chosen to obtain $T_{\rm e} \text{ ([S\thinspace III])}= T_{\rm e} \text{ ([O\thinspace III])}$. The result for the northern knot is independent.}, while the modification implies $T_{\rm e} \text{ ([S\thinspace III])}\approx T_{\rm e} \text{ ([N\thinspace II])}$ for the Orion Nebula. The latter would imply that the S$^{2+}$ and N$^{+}$ ions coexist almost exclusively in the zone of low degree of ionization. However, the large increment we have assumed in the collision strengths are unable to give a \textit{normal}\footnote{A value consistent with that of the Orion Nebula.} S abundance in the high-velocity components. If we adopt $\text{12+log(S/H)}=7.07 \pm 0.03$ from \citet{mendez2021-2} as reference, we still have an overestimation of 0.23 dex in the jet base. A similar difference is obtained in the knot. Moreover, this experiment produces a $\text{log(S/O)} = -1.81 \pm 0.06$ for the nebular component, which is inconsistent with the solar value of $\text{log(S/O)}=-1.58 \pm 0.08$ \citep{lodders19}. Since both S and O are $\alpha$-elements, a constant abundance ratio is expected from their common nucleosynthesis. Therefore, following a \textit{reductio ad absurdum} process, we can say that large errors in the collision strengths $\Omega_{9531}$, $\Omega_{9069}$ and $\Omega_{8829}$ do not solve the observed discrepancies in  $T_{\rm e} \text{ ([S\thinspace III])}$ and the total S abundance in HH~514.

\subsubsection{Is the [S\thinspace III] emission affected by the shock?}
\label{subsec:shock_affected}

HH objects are produced by the interaction of high-velocity flows
with the surrounding medium or by internal shocks within the flow.
In both cases, gas is initially heated and compressed in a shock front,
followed by a non-equilibrium cooling zone in which the gas returns to the
photoionization equilibrium temperature while undergoing further compression.
The cooling zone may contribute to observed CEL spectrum,
but the greatest effect is on highly ionized lines that are not typically seen
in \hii{} regions.

In both the jet and knot of HH~514 the low observed line widths of about
\SI{20}{km.s^{-1}}
(see Table~\ref{tab:sample_spectra})
imply that only low velocity internal shocks are present.
For moderate ionization lines such as [\ion{S}{3}], the relative contribution is therefore expected to be small \citep{mendez2021}.
Moreover, the emission from the cooling zone, if any, would affect $T_{\rm e} \text{ ([S\thinspace III])}$ in the opposite direction, increasing the observed [S\thinspace III] $I(\lambda 6312)/I(\lambda 9531)$ line intensity ratio that would produce higher temperatures.   

\subsubsection{Telluric absorption bands or sky emission?}
\label{subsec:tell_abs_sky_emmision}

There are no other lines arising from  sky emission or the Orion Nebula that could be blended with [S\thinspace III] $\lambda \lambda 6312, 9069, 9531$ ($\lambda \lambda 6315.31,9073.60, 9535.87$) lines at the spectral resolution of our spectra. As commented in Sect.~\ref{subsec:atomic_data}, we correct for telluric absorptions that could affect the intensity of [S\thinspace III] $\lambda \lambda 9531, 9069$ lines in our spectra. Moreover, those [S\thinspace III] lines fall outside the absorption bands in  the spectra of HH~514. In any case, in the presence of absorption bands, their effect on [S\thinspace III] $\lambda \lambda 9531, 9069$ lines would fictitiously increase $T_{\rm e} \text{ ([S\thinspace III])}$ and decrease the S$^{2+}$ abundance, which is the opposite of what is found.

\subsubsection{Errors in the reddening correction?}
\label{subsec:errors_CHB}

The [S\thinspace III] $I(\lambda 6312)/I(\lambda 9531)$ line intensity ratio is highly dependent on the reddening correction. However, if we estimate what reddening coefficient is necessary in order to produce $T_{\rm e} \text{ ([S\thinspace III])} \geq T_{\rm e} \text{ ([O\thinspace III])}$, we get $\text{c}(\text{H}\beta)\geq 1.22$. This is at least 44 per cent higher than the estimated value based on the observed Balmer and Paschen H\thinspace I lines, which is simply not possible considering the quality of the flux calibration of our data. There is also no evidence of such an increase in c(H$\beta$) in the MUSE data\citep[][]{Weilbacher15}, which is consistent with our results. In any case, the derivation of a largely incorrect $\text{c}(\text{H}\beta)$ value would produce many different and noticeable inconsistencies in our results that are obviously not observed. 

\subsubsection{Is the density  overestimated?}
\label{subsec:ov_density}

Let us suppose that the representative density of the [S\thinspace III] zone is overestimated and its true value is more similar to that of the Orion Nebula. As can be seen in panel (a) of Fig.~\ref{fig:plasma}, this would imply that in the jet base, $T_{\rm e} \text{ ([S\thinspace III])} \approx 10200 \text{ K}$ if $n_{\rm e} \leq 10000 \text{ cm}^{-3}$. However, even assuming that value of $T_{\rm e} \text{ ([S\thinspace III])}$, the S$^{2+}$ abundance would be still overestimated. A $T_{\rm e} \text{ ([S\thinspace III])} \approx 13000 \text{ K}$ would be necessary to obtain a {\it normal} S$^{2+}$ abundance, which can not be achieved with any density value.

\subsection{Comparison between HH~514 and its associated proplyd}
\label{subsec:proplyd}

In the previous section, we established that none of the potential systematic biases
can plausibly cause the observed \siii{} emission from the HH~514 jet,
leaving an enhanced sulfur abundance as the most likely explanation.
Given that the ionized base of the jet emerges from the ionization front
of the proplyd 170-337 \citep{bally00},
it is important to compare the abundances in the two components.

\begin{figure}
\includegraphics[width=\columnwidth]{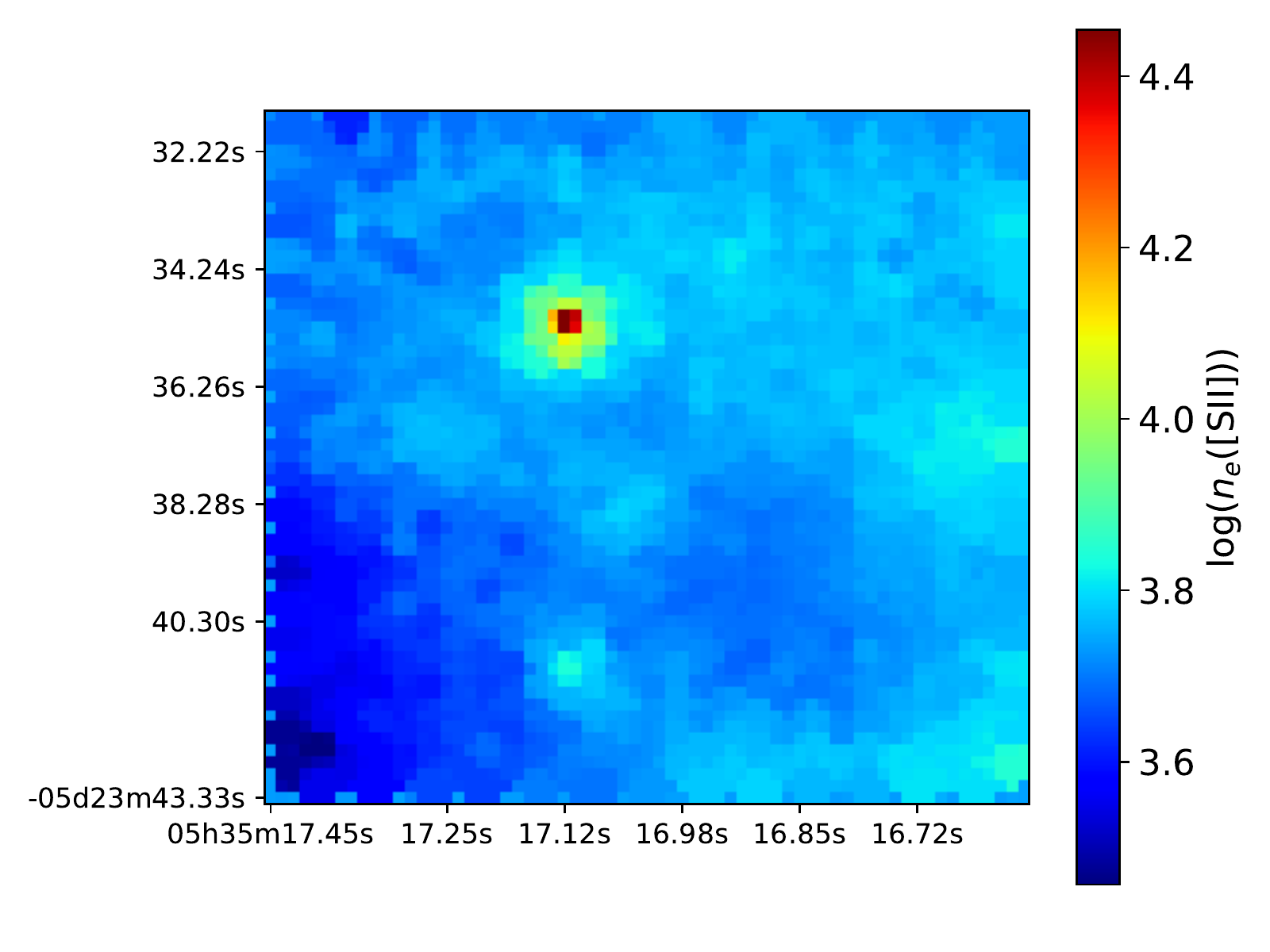}
\caption{Density map obtained with the intensity ratio [S\thinspace II] $\lambda$6731/$\lambda$6716 using the MEGARA-IFU observations (see Fig.~\ref{fig:megara}). This map reveals in a partial way only, the density structures present in the Field of View (see text).}
\label{fig:density_map}
\end{figure}

In our UVES observations, 170-337 is within Cut~1. We interpolate the nebular emission above and below the zone covered by 170-337 along the slit, so that we can separate the proplyd emission from that of the Orion Nebula.
The line intensities used are shown in Table~\ref{tab:fluxes_proplyd}.
The complete study of 170-337 will be analyzed in a future paper (M\'endez-Delgado et al., in prep).
By analyzing the spectrum, we obtain  $n_{\rm e}\approx 10^{5.85 \pm 0.20} \text{ cm}^{-3}$ from the $n_{\rm e}$([S\thinspace II]) $\lambda$4069/$\lambda$4076 diagnostic (no [Fe\thinspace III] emission lines are detected in 170-337). We use those [S\thinspace II] lines -- that originate from the upper levels \chem{^2P} -- because of their very high critical densities, above the $n_{\rm e}$ value obtained for the proplyd.
This density is close to the value of \SI{1.3e6}{cm^{-3}}
found by \citet{henney99} for the same proplyd
from modelling the H\(\alpha\) surface brightness profile. 
In Fig.~\ref{fig:density_map} we show the 2-D density map obtained with MEGARA from the [S\thinspace II] $\lambda$6731/$\lambda$6716 ratio. Lines that come from the intermediate \chem{^2D} levels of \chem{S^{+}} and have a considerable lower critical density. In Fig.~\ref{fig:density_map}, we can see that 170-337 shows a rather low contrast with respect to the surroundings, because the emission of [S\thinspace II] $\lambda\lambda$6731, 6716 lines is dominated by the background much lower-density nebular emission. High-density clumps as the proplyd 170-337 can contribute appreciably to the emission of lines arising from the higher levels --as the $T_{\rm e}$-sensitive auroral lines--, which are less affected by collisional deexcitation. The use of incorrect density diagnostics can lead to wrong estimates of physical conditions and chemical abundances \citep[][]{mendez2021-2}. 

By adopting the correct density for the ionized layers of 170-337, we obtain $T_{\rm e} \text{ ([O\thinspace III])} \approx T_{\rm e} \text{ ([S\thinspace III])} \approx T_{\rm e} \text{ ([N\thinspace II])} \approx 8000 \pm 1000 \text{ K}$, in total consistency with the expected values for the gas in photoionization equilibrium within the radiation field of the Orion Trapezium. Although the uncertainties associated with the three temperature diagnostics are high, due to the extremely large density of the proplyd, the convergence of the diagnostics ensures the reliability of the derived physical conditions, yielding the adopted temperature $T_{\rm e}=\SI{8010 \pm 60}{K}$. With the aforementioned physical conditions we obtain $\text{12+log(S/H)} \approx 7.10 \pm 0.04 $ which is consistent with the value derived by \citet{mendez2021-2} for HH~204, representative of the chemical composition of the Orion Nebula. Therefore, the photoionized material of the outer layers of proplyd 170-337 does not show an overabundance of S.

\subsection{Implications of the elevated S abundance in HH~514}
\label{subsec:real_overabundance}

From section~\ref{subsec:under_TS3}, we have ruled out several scenarios that may produce an unrealistic overabundance of S and a low $T_{\rm e} \text{ ([S\thinspace III])}$.
Therefore, it is possible to affirm that the overabundance of S/H in HH~514 is real.
Furthermore, neither the surrounding nebula nor the proplyd photoevaporation flow
show the same overabundance.
The other significant difference between the gas phase abundances in the
two flows is the approximately solar Fe and Ni abundance in HH~514,
compared with severely depleted abundances in the proplyd (similar to the nebula).
This is most simply explained by dust destruction in the HH object,
which liberates Fe and Ni to the gas phase, while it remains locked within grains in the case of the proplyd.
It is therefore reasonable to suppose that the difference in S abundances
might also be related to processes involving dust grains.

It is common to observe relatively high gas-phase Fe and Ni abundances
in HH objects, both in photoionized
\citep[][]{Blagrave06, mesadelgado09, mendez2021, mendez2021-2}
and neutral \citep{Nisini:2002f, Giannini:2015s}
environments.
Depletions with respect to solar abundance of zero to \SI{0.3}{dex} are typically found,
as compared to a depletion of more than \SI{1}{dex} in the general ISM and \hii{} regions
\citep{Rodriguez:2002a, Jenkins09}.
This is usually interpreted as evidence for dust reprocessing in the HH shocks themselves,
although there is also evidence that some reprocessing must have occurred in
the jet launch region \citep{Bohm:2001a, Podio:2009a}.
However, the present study is the first to find a super-solar sulphur abundance in an HH object.

It is well known that planetary formation process is capable of building up large grains in protoplanetary disks \citep[][]{Beckwith2000}.
Recently, \citet{Kama19} have suggested the formation of large sulfide mineral reservoirs during the accretion process from the protoplanetary disk to a central star.
The presence of a giant planet immersed in the disk, can cause a pressure bump that can accumulate the largest dust particles \citep{Pinilla2012, Birnstiel16}, among which would be the sulfides.
If some mechanism allowed these sulfide grains to preferentially enter the jet flow from the inner disk,
and that the grains could subsequently be destroyed,
then this could explain the high sulphur abundances that we observe in HH~514.
The exact mechanism by which this might be accomplished is unclear however.
Simulations of disk winds that include the coupling of different-sized grains
to the flow have recently been developed \citep{Rodenkirch:2022o},
but these do not yet include the effects of external photoevaporation,
nor do they include the wind from the innermost disk that gives rise to the fast jet.

\subsection{The total abundances of Fe and Ni}
\label{subsec:totalFeandNi}

In the jet base component of HH~514 we are able to derive the abundances of Ni$^{+}$, Ni$^{2+}$, Fe$^{+}$ and Fe$^{2+}$. As mentioned in Sec.~\ref{sec:ionic_total_abundances}, despite only having an upper limit to the abundance of Fe$^{3+}$, the ICF schemes of \citet{rodriguez05} indicate that the total abundance of Fe/H should be close to the sum of the abundances of Fe$^{+}$ and Fe$^{2+}$ and the upper limit of Fe$^{3+}$. Both the total abundances of Ni and Fe in the jet base are a factor 34 higher than in the main emission from the Orion Nebula. Moreover, these values are very close to the reference Solar values from \citet{lodders19}. However, this does not necessarily mean that all the dust grains containing Fe and Ni have been destroyed. This is also supported by the evidence of thermal dust IR emission from HH~514 \citep{smith05}, which suggests the survival of some amount of dust in the object. In connection with Sec.~\ref{subsec:real_overabundance}, HH~514 is able to destroy the dust reservoirs of S in a planet-forming disk. However, it can also destroy dust grains of zones of the Orion Nebula outside the S reservoirs, with normal S abundance. To estimate the fraction of Fe released from each kind of dust grains, we can assume that the S excess is produced by the destruction of FeS compounds, as suggested by \citet{Kama19}. In addition, NiS compounds should be present as well. For each atom of S released, one atom of Fe or Ni should be released too. Since the S excess will be equal to the difference between the S abundance in HH~514 and the representative value of the Orion Nebula, then:
\begin{equation*}
    \frac{n(\text{S})_{\text{excess}}}{n(\text{H})}\approx\frac{n(\text{Fe+Ni})_{\text{excess}}}{n(\text{H})}\approx\frac{1.05\times n(\text{Fe})_{\text{excess}}}{n(\text{H})}=1.52\times 10^{-5},
\end{equation*}
adopting $\text{12+log(S/H)}=7.07 \pm 0.03$ \citep{mendez2021-2} for the Orion Nebula. Considering the Fe and Ni abundances obtained in the nebular components, this would imply that of the total abundances of Fe/H and Ni/H in the jet base, $\sim$67 per cent would come from dust destruction of S reservoirs, $\sim$30 per cent from the destruction of dust grains of normal S abundance and $\sim$3 per cent from Fe and Ni that were already in the gaseous phase. 



\section{Conclusions}
\label{sec:summary_and_conclusions}
We analyze the physical conditions and chemical abundances of the photoionized Herbig-Haro object HH~514 and the Orion Nebula making use of high-spectral resolution spectra taken with UVES at VLT, high-spatial resolution imaging from the HST and deep MEGARA-IFU spectra.  
Due to the Doppler shift between the emission of the Orion Nebula and HH~514, we independently study the emission of two high-velocity components: the jet base and a knot. These HHs are extremely dense: $n_{\rm e}(\text{HH~514~jet base})=10^{5.37} \text{ cm}^{-3}$ and $n_{\rm e}(\text{HH~514~knot})=10^{4.87} \text{ cm}^{-3}$. However, they present temperatures consistent with  photoionization equilibrium in the Orion Nebula, $T_{\rm e} \text{ ([O\thinspace III])} \sim 9000\text{ K}$.

The total abundances of the high-velocity components are fully consistent with the ones derived in the Orion Nebula with the exception of S, Fe, and Ni. In the case of S, the S$^{2+}$ ion comprises 91 per cent of the total abundance.
Furthermore, we found that $T_{\rm e} \text{ ([S\thinspace III])} < T_{\rm e} \text{ ([O\thinspace III])} $, which is difficult to explain under normal photoionization conditions \citep[][]{Binette2012}.
We rule out the possibility that the S overabundance may be a spurious consequence
of errors in the atomic data or in the observations.
We speculate that it may be due to the preferential incorporation of S-rich dust into
the high-velocity jet from the inner disk, which is then returned to the gas phase
either in the jet launching region or via internal shocks in the jet.
The origin of S-rich dust reservoirs maybe related to the process
that \citet{Kama19} propose to occur in protoplanetary discs,
which consists of the accumulation of large sulfide dust grains
around giant planets during accretion from the disc to the central star.

In the jet base we obtain log(Fe/H)+12$ =  7.33 \pm 0.02$ and log(Ni/H)+12$ = 6.03 \pm 0.08$. These abundances are a factor 34 higher than those derived in the Orion Nebula and are very close to the Solar values \citep[][]{lodders19}.
We estimate that $\sim$67 per cent of the total Fe/H and Ni/H abundances in the jet base have origin in the destruction of sulfide reservoirs, whereas $\sim$30 per cent would come from the destruction of dust grains with a poorer S content, having just $\sim$3 per cent from Fe and Ni previously present in the gaseous phase.
HH~514 destroys dust grains of the Orion Nebula gas more efficiently than other HHs studied in the literature such as HH~202~S, HH~529~II, HH~529~III and HH~204 \citep[][]{mesadelgado09, mendez2021, mendez2021-2}.

Knowing the relationship between the efficiency of dust destruction and the propagation velocity of the HH object is an important issue. In our series of papers about the chemical content of HH objects, we find that gas-phase Fe/H and Ni/H abundances seem to be correlated with the propagation velocity (being HH~529 the slowest and HH~514 the fastest), which is consistent with studies of HH objects propagating in neutral environments \citep[See Fig.~14 from][]{Hartigan2020}. However, it is still not clear if this relation is simply linear or shows a more complex behaviour or even if the dust destruction efficiency reaches a maximum at a given velocity. Some parameters such as the degree of ionization may also play a role. The amount of data is still too limited for addressing this question and analyses of further HH objects are still needed.

Comparison of the optical jet kinematics with the kinematics of a CO molecular emission line
close to the central star yields contradictory results.
If the molecular line traces the disk rotation, then the jet is perpendicular to the rotation axis,
but if the molecular line traces outflow, then the outflow is perpendicular to the jet.
We suggest instead that the molecular line may trace inflowing material
that is the remnant of the cloud core from which the star formed.

\section*{Acknowledgements}
We are grateful to the anonymous referee for his/her helpful comments. This work is based on observations collected at the European Southern Observatory, Chile, proposal number ESO 092.C-0323(A), and on observations made with the Gran Telescopio Canarias (GTC), installed in the Spanish Observatorio del Roque de los Muchachos of the Instituto de Astrofísica de Canarias, in the island of La Palma. The authors appreciate the friendly communication of Dr. Richard Booth on the planet formation processes. We acknowledge support from the Agencia Estatal de Investigaci\'on del Ministerio de Ciencia e Innovaci\'on (AEI-MCINN) under grant {\it Espectroscop\'ia de campo integral de regiones \ion{H}{2} locales. Modelos para el estudio de regiones \ion{H}{2} extragal\'acticas} with reference 10.13039/501100011033.  JEM-D thanks the support of the Instituto de Astrof\'isica de Canarias under the Astrophysicist Resident Program and acknowledges support from the Mexican CONACyT (grant CVU 602402). JG-R acknowledges financial support from the Spanish Ministry of Science and Innovation (MICINN) through the Spanish State Research Agency, under Severo Ochoa Centres of Excellence Programme 2020-2023 (CEX2019-000920-S), and from the Canarian Agency for Research, Innovation and Information Society (ACIISI), of the Canary Islands Government, and the European Regional Development Fund (ERDF), under grant with reference ProID2021010074. The authors acknowledge support under grant P/308614 financed by funds transferred from the Spanish Ministry of Science, Innovation and Universities, charged to the General State Budgets and with funds transferred from the General Budgets of the Autonomous Community of the Canary Islands by the MCIU. The paper has been edited using the Overleaf facility.

\section*{DATA AVAILABILITY}
The lines measured in the spectra are entirely available in the supplementary material of this article. Table~\ref{tab:sample_spectra} is an example of its content. The rest of information is found in tables or references of this paper.



\bibliographystyle{mnras}
\bibliography{Mendez}

\begin{thebibliography}{}
\makeatletter
\relax
\def\mn@urlcharsother{\let\do\@makeother \do\$\do\&\do\#\do\^\do\_\do\%\do\~}
\def\mn@doi{\begingroup\mn@urlcharsother \@ifnextchar [ {\mn@doi@}
  {\mn@doi@[]}}
\def\mn@doi@[#1]#2{\def\@tempa{#1}\ifx\@tempa\@empty \href
  {http://dx.doi.org/#2} {doi:#2}\else \href {http://dx.doi.org/#2} {#1}\fi
  \endgroup}
\def\mn@eprint#1#2{\mn@eprint@#1:#2::\@nil}
\def\mn@eprint@arXiv#1{\href {http://arxiv.org/abs/#1} {{\tt arXiv:#1}}}
\def\mn@eprint@dblp#1{\href {http://dblp.uni-trier.de/rec/bibtex/#1.xml}
  {dblp:#1}}
\def\mn@eprint@#1:#2:#3:#4\@nil{\def\@tempa {#1}\def\@tempb {#2}\def\@tempc
  {#3}\ifx \@tempc \@empty \let \@tempc \@tempb \let \@tempb \@tempa \fi \ifx
  \@tempb \@empty \def\@tempb {arXiv}\fi \@ifundefined
  {mn@eprint@\@tempb}{\@tempb:\@tempc}{\expandafter \expandafter \csname
  mn@eprint@\@tempb\endcsname \expandafter{\@tempc}}}

\bibitem[\protect\citeauthoryear{{Adams}, {Hollenbach}, {Laughlin}  \&
  {Gorti}}{{Adams} et~al.}{2004}]{Adams:2004u}
{Adams} F.~C.,  {Hollenbach} D.,  {Laughlin} G.,   {Gorti} U.,  2004, \mn@doi
  [\apj] {10.1086/421989}, \href
  {https://ui.adsabs.harvard.edu/abs/2004ApJ...611..360A} {611, 360}

\bibitem[\protect\citeauthoryear{{Amayo}, {Delgado-Inglada}  \&
  {Stasi{\'n}ska}}{{Amayo} et~al.}{2021}]{Amayo2021}
{Amayo} A.,  {Delgado-Inglada} G.,   {Stasi{\'n}ska} G.,  2021, \mn@doi
  [\mnras] {10.1093/mnras/stab1467}, \href
  {https://ui.adsabs.harvard.edu/abs/2021MNRAS.505.2361A} {505, 2361}

\bibitem[\protect\citeauthoryear{{Baldwin} et~al.,}{{Baldwin}
  et~al.}{1996}]{Baldwin96}
{Baldwin} J.~A.,  et~al., 1996, \mn@doi [\apjl] {10.1086/310245}, \href
  {https://ui.adsabs.harvard.edu/abs/1996ApJ...468L.115B} {468, L115}

\bibitem[\protect\citeauthoryear{{Bally}, {Sutherland}, {Devine}  \&
  {Johnstone}}{{Bally} et~al.}{1998}]{Bally:1998a}
{Bally} J.,  {Sutherland} R.~S.,  {Devine} D.,   {Johnstone} D.,  1998, \mn@doi
  [\aj] {10.1086/300399}, \href
  {http://adsabs.harvard.edu/abs/1998AJ....116..293B} {116, 293}

\bibitem[\protect\citeauthoryear{{Bally}, {O'Dell}  \& {McCaughrean}}{{Bally}
  et~al.}{2000}]{bally00}
{Bally} J.,  {O'Dell} C.~R.,   {McCaughrean} M.~J.,  2000, \mn@doi [\aj]
  {10.1086/301385}, \href
  {https://ui.adsabs.harvard.edu/abs/2000AJ....119.2919B} {119, 2919}

\bibitem[\protect\citeauthoryear{{Bally}, {Licht}, {Smith}  \&
  {Walawender}}{{Bally} et~al.}{2006}]{Bally:2006a}
{Bally} J.,  {Licht} D.,  {Smith} N.,   {Walawender} J.,  2006, \mn@doi [\aj]
  {10.1086/498265}, \href {http://adsabs.harvard.edu/abs/2006AJ....131..473B}
  {131, 473}

\bibitem[\protect\citeauthoryear{{Bautista}, {Peng}  \& {Pradhan}}{{Bautista}
  et~al.}{1996}]{Bautista96}
{Bautista} M.~A.,  {Peng} J.,   {Pradhan} A.~K.,  1996, \mn@doi [\apj]
  {10.1086/176976}, \href
  {https://ui.adsabs.harvard.edu/abs/1996ApJ...460..372B} {460, 372}

\bibitem[\protect\citeauthoryear{{Beckwith}, {Henning}  \&
  {Nakagawa}}{{Beckwith} et~al.}{2000}]{Beckwith2000}
{Beckwith} S.~V.~W.,  {Henning} T.,   {Nakagawa} Y.,  2000, in {Mannings} V.,
  {Boss} A.~P.,   {Russell} S.~S.,  eds, Protostars and Planets IV. p.~533
  (\mn@eprint {arXiv} {astro-ph/9902241})

\bibitem[\protect\citeauthoryear{{Berg}, {Pogge}, {Skillman}, {Croxall},
  {Moustakas}, {Rogers}  \& {Sun}}{{Berg} et~al.}{2020}]{Berg20}
{Berg} D.~A.,  {Pogge} R.~W.,  {Skillman} E.~D.,  {Croxall} K.~V.,  {Moustakas}
  J.,  {Rogers} N. S.~J.,   {Sun} J.,  2020, \mn@doi [\apj]
  {10.3847/1538-4357/ab7eab}, \href
  {https://ui.adsabs.harvard.edu/abs/2020ApJ...893...96B} {893, 96}

\bibitem[\protect\citeauthoryear{{Binder} \& {Povich}}{{Binder} \&
  {Povich}}{2018}]{Binder2018}
{Binder} B.~A.,  {Povich} M.~S.,  2018, \mn@doi [\apj]
  {10.3847/1538-4357/aad7b2}, \href
  {https://ui.adsabs.harvard.edu/abs/2018ApJ...864..136B} {864, 136}

\bibitem[\protect\citeauthoryear{{Binette}, {Matadamas}, {H{\"a}gele},
  {Nicholls}, {Magris C.}, {Pe{\~n}a-Guerrero}, {Morisset}  \&
  {Rodr{\'\i}guez-Gonz{\'a}lez}}{{Binette} et~al.}{2012}]{Binette2012}
{Binette} L.,  {Matadamas} R.,  {H{\"a}gele} G.~F.,  {Nicholls} D.~C.,  {Magris
  C.} G.,  {Pe{\~n}a-Guerrero} M.~{\'A}.,  {Morisset} C.,
  {Rodr{\'\i}guez-Gonz{\'a}lez} A.,  2012, \mn@doi [\aap]
  {10.1051/0004-6361/201219515}, \href
  {https://ui.adsabs.harvard.edu/abs/2012A&A...547A..29B} {547, A29}

\bibitem[\protect\citeauthoryear{{Birnstiel}, {Fang}  \&
  {Johansen}}{{Birnstiel} et~al.}{2016}]{Birnstiel16}
{Birnstiel} T.,  {Fang} M.,   {Johansen} A.,  2016, \mn@doi [\ssr]
  {10.1007/s11214-016-0256-1}, \href
  {https://ui.adsabs.harvard.edu/abs/2016SSRv..205...41B} {205, 41}

\bibitem[\protect\citeauthoryear{{Blagrave}, {Martin}  \& {Baldwin}}{{Blagrave}
  et~al.}{2006}]{Blagrave06}
{Blagrave} K.~P.~M.,  {Martin} P.~G.,   {Baldwin} J.~A.,  2006, \mn@doi [\apj]
  {10.1086/503830}, \href
  {https://ui.adsabs.harvard.edu/abs/2006ApJ...644.1006B} {644, 1006}

\bibitem[\protect\citeauthoryear{{Blagrave}, {Martin}, {Rubin}, {Dufour},
  {Baldwin}, {Hester}  \& {Walter}}{{Blagrave} et~al.}{2007}]{Blagrave07}
{Blagrave} K.~P.~M.,  {Martin} P.~G.,  {Rubin} R.~H.,  {Dufour} R.~J.,
  {Baldwin} J.~A.,  {Hester} J.~J.,   {Walter} D.~K.,  2007, \mn@doi [\apj]
  {10.1086/510151}, \href
  {https://ui.adsabs.harvard.edu/abs/2007ApJ...655..299B} {655, 299}

\bibitem[\protect\citeauthoryear{{Boyden} \& {Eisner}}{{Boyden} \&
  {Eisner}}{2020}]{Boyden:2020a}
{Boyden} R.~D.,  {Eisner} J.~A.,  2020, \mn@doi [\apj]
  {10.3847/1538-4357/ab86b7}, \href
  {https://ui.adsabs.harvard.edu/abs/2020ApJ...894...74B} {894, 74}

\bibitem[\protect\citeauthoryear{{Böhm} \& {Matt}}{{Böhm} \&
  {Matt}}{2001}]{Bohm:2001a}
{Böhm} K.-H.,  {Matt} S.,  2001, \mn@doi [\pasp] {10.1086/318611}, \href
  {https://ui.adsabs.harvard.edu/abs/2001PASP..113..158B} {113, 158}

\bibitem[\protect\citeauthoryear{{D'Odorico}, {Cristiani}, {Dekker}, {Hill},
  {Kaufer}, {Kim}  \& {Primas}}{{D'Odorico} et~al.}{2000}]{Dodorico00}
{D'Odorico} S.,  {Cristiani} S.,  {Dekker} H.,  {Hill} V.,  {Kaufer} A.,  {Kim}
  T.,   {Primas} F.,  2000, {Performance of UVES, the echelle spectrograph for
  the ESO VLT and highlights of the first observations of stars and quasars}.
pp 121--130, \mn@doi{10.1117/12.390133}

\bibitem[\protect\citeauthoryear{{Del Zanna}, {Dere}, {Young}  \& {Landi}}{{Del
  Zanna} et~al.}{2021}]{Chianti10}
{Del Zanna} G.,  {Dere} K.~P.,  {Young} P.~R.,   {Landi} E.,  2021, \mn@doi
  [\apj] {10.3847/1538-4357/abd8ce}, \href
  {https://ui.adsabs.harvard.edu/abs/2021ApJ...909...38D} {909, 38}

\bibitem[\protect\citeauthoryear{{Duchêne}, {Lacour}, {Moraux}, {Goodwin}  \&
  {Bouvier}}{{Duchêne} et~al.}{2018}]{Duchene:2018a}
{Duchêne} G.,  {Lacour} S.,  {Moraux} E.,  {Goodwin} S.,   {Bouvier} J.,
  2018, \mn@doi [\mnras] {10.1093/mnras/sty1180}, \href
  {https://ui.adsabs.harvard.edu/abs/2018MNRAS.478.1825D} {478, 1825}

\bibitem[\protect\citeauthoryear{{Eisner} et~al.,}{{Eisner}
  et~al.}{2018}]{Eisner:2018a}
{Eisner} J.~A.,  et~al., 2018, \mn@doi [\apj] {10.3847/1538-4357/aac3e2}, \href
  {http://adsabs.harvard.edu/abs/2018ApJ...860...77E} {860, 77}

\bibitem[\protect\citeauthoryear{{Esteban}, {Peimbert}, {Garc{\'\i}a-Rojas},
  {Ruiz}, {Peimbert}  \& {Rodr{\'\i}guez}}{{Esteban} et~al.}{2004}]{Esteban04}
{Esteban} C.,  {Peimbert} M.,  {Garc{\'\i}a-Rojas} J.,  {Ruiz} M.~T.,
  {Peimbert} A.,   {Rodr{\'\i}guez} M.,  2004, \mn@doi [\mnras]
  {10.1111/j.1365-2966.2004.08313.x}, \href
  {https://ui.adsabs.harvard.edu/abs/2004MNRAS.355..229E} {355, 229}

\bibitem[\protect\citeauthoryear{{Esteban}, {Carigi}, {Copetti},
  {Garc{\'\i}a-Rojas}, {Mesa-Delgado}, {Casta{\~n}eda}  \&
  {P{\'e}quignot}}{{Esteban} et~al.}{2013}]{Esteban13}
{Esteban} C.,  {Carigi} L.,  {Copetti} M.~V.~F.,  {Garc{\'\i}a-Rojas} J.,
  {Mesa-Delgado} A.,  {Casta{\~n}eda} H.~O.,   {P{\'e}quignot} D.,  2013,
  \mn@doi [\mnras] {10.1093/mnras/stt730}, \href
  {https://ui.adsabs.harvard.edu/abs/2013MNRAS.433..382E} {433, 382}

\bibitem[\protect\citeauthoryear{{Froese Fischer}, {Tachiev}  \&
  {Irimia}}{{Froese Fischer} et~al.}{2006}]{FFTI06}
{Froese Fischer} C.,  {Tachiev} G.,   {Irimia} A.,  2006, \mn@doi [Atomic Data
  and Nuclear Data Tables] {10.1016/j.adt.2006.03.001}, \href
  {https://ui.adsabs.harvard.edu/abs/2006ADNDT..92..607F} {92, 607}

\bibitem[\protect\citeauthoryear{{Galavis}, {Mendoza}  \& {Zeippen}}{{Galavis}
  et~al.}{1995}]{GMZ95}
{Galavis} M.~E.,  {Mendoza} C.,   {Zeippen} C.~J.,  1995, \aaps, \href
  {https://ui.adsabs.harvard.edu/abs/1995A&AS..111..347G} {111, 347}

\bibitem[\protect\citeauthoryear{{Garc{\'\i}a-Rojas}, {Esteban}, {Peimbert},
  {Rodr{\'\i}guez}, {Ruiz}  \& {Peimbert}}{{Garc{\'\i}a-Rojas}
  et~al.}{2004}]{garciarojas04}
{Garc{\'\i}a-Rojas} J.,  {Esteban} C.,  {Peimbert} M.,  {Rodr{\'\i}guez} M.,
  {Ruiz} M.~T.,   {Peimbert} A.,  2004, \mn@doi [\apjs] {10.1086/421909}, \href
  {https://ui.adsabs.harvard.edu/abs/2004ApJS..153..501G} {153, 501}

\bibitem[\protect\citeauthoryear{{Garc{\'\i}a-Rojas}, {Esteban}, {Peimbert},
  {Peimbert}, {Rodr{\'\i}guez}  \& {Ruiz}}{{Garc{\'\i}a-Rojas}
  et~al.}{2005}]{garciarojas05}
{Garc{\'\i}a-Rojas} J.,  {Esteban} C.,  {Peimbert} A.,  {Peimbert} M.,
  {Rodr{\'\i}guez} M.,   {Ruiz} M.~T.,  2005, \mn@doi [\mnras]
  {10.1111/j.1365-2966.2005.09302.x}, \href
  {https://ui.adsabs.harvard.edu/abs/2005MNRAS.362..301G} {362, 301}

\bibitem[\protect\citeauthoryear{{Garc{\'\i}a-Rojas}, {Esteban}, {Peimbert},
  {Costado}, {Rodr{\'\i}guez}, {Peimbert}  \& {Ruiz}}{{Garc{\'\i}a-Rojas}
  et~al.}{2006}]{garciarojas06}
{Garc{\'\i}a-Rojas} J.,  {Esteban} C.,  {Peimbert} M.,  {Costado} M.~T.,
  {Rodr{\'\i}guez} M.,  {Peimbert} A.,   {Ruiz} M.~T.,  2006, \mn@doi [\mnras]
  {10.1111/j.1365-2966.2006.10105.x}, \href
  {https://ui.adsabs.harvard.edu/abs/2006MNRAS.368..253G} {368, 253}

\bibitem[\protect\citeauthoryear{{Garc{\'\i}a-Rojas}, {Esteban}, {Peimbert},
  {Rodr{\'\i}guez}, {Peimbert}  \& {Ruiz}}{{Garc{\'\i}a-Rojas}
  et~al.}{2007}]{garciarojas07-2}
{Garc{\'\i}a-Rojas} J.,  {Esteban} C.,  {Peimbert} A.,  {Rodr{\'\i}guez} M.,
  {Peimbert} M.,   {Ruiz} M.~T.,  2007, \rmxaa, \href
  {https://ui.adsabs.harvard.edu/abs/2007RMxAA..43....3G} {43, 3}

\bibitem[\protect\citeauthoryear{{Giannini}, {Antoniucci}, {Nisini},
  {Bacciotti}  \& {Podio}}{{Giannini} et~al.}{2015}]{Giannini:2015s}
{Giannini} T.,  {Antoniucci} S.,  {Nisini} B.,  {Bacciotti} F.,   {Podio} L.,
  2015, \mn@doi [\apj] {10.1088/0004-637X/814/1/52}, \href
  {https://ui.adsabs.harvard.edu/abs/2015ApJ...814...52G} {814, 52}

\bibitem[\protect\citeauthoryear{{Gil De Paz}, {Pascual}  \&
  {Chamorro-Cazorla}}{{Gil De Paz} et~al.}{2020}]{gildepaz+20}
{Gil De Paz} A.,  {Pascual} S.,   {Chamorro-Cazorla} M.,  2020,
  {guaix-ucm/megara-tools: Release v0.1.1}, \mn@doi{10.5281/zenodo.4264048}

\bibitem[\protect\citeauthoryear{{Gil de Paz} et~al.,}{{Gil de Paz}
  et~al.}{2018}]{gildepaz+18}
{Gil de Paz} A.,  et~al., 2018, in {Evans} C.~J.,  {Simard} L.,   {Takami} H.,
  eds,  Society of Photo-Optical Instrumentation Engineers (SPIE) Conference
  Series Vol. 10702, Ground-based and Airborne Instrumentation for Astronomy
  VII. p. 1070217, \mn@doi{10.1117/12.2313299}

\bibitem[\protect\citeauthoryear{{Grieve}, {Ramsbottom}, {Hudson}  \&
  {Keenan}}{{Grieve} et~al.}{2014}]{GRHK14}
{Grieve} M.~F.~R.,  {Ramsbottom} C.~A.,  {Hudson} C.~E.,   {Keenan} F.~P.,
  2014, \mn@doi [\apj] {10.1088/0004-637X/780/1/110}, \href
  {https://ui.adsabs.harvard.edu/abs/2014ApJ...780..110G} {780, 110}

\bibitem[\protect\citeauthoryear{{Hanuschik}}{{Hanuschik}}{2003}]{Hanuschik03}
{Hanuschik} R.~W.,  2003, \mn@doi [\aap] {10.1051/0004-6361:20030885}, \href
  {https://ui.adsabs.harvard.edu/abs/2003A&A...407.1157H} {407, 1157}

\bibitem[\protect\citeauthoryear{{Hartigan}, {Hillenbrand}, {Matuszewski},
  {Borges}, {Neill}, {Martin}, {Morrissey}  \& {Moore}}{{Hartigan}
  et~al.}{2020}]{Hartigan2020}
{Hartigan} P.,  {Hillenbrand} L.~A.,  {Matuszewski} M.,  {Borges} A.~C.,
  {Neill} J.~D.,  {Martin} D.~C.,  {Morrissey} P.,   {Moore} A.~M.,  2020,
  \mn@doi [\aj] {10.3847/1538-3881/abadfa}, \href
  {https://ui.adsabs.harvard.edu/abs/2020AJ....160..165H} {160, 165}

\bibitem[\protect\citeauthoryear{{Hasegawa} et~al.,}{{Hasegawa}
  et~al.}{2022}]{Hasegawa:2022j}
{Hasegawa} Y.,  et~al., 2022, \mn@doi [\apjl] {10.3847/2041-8213/ac50aa}, \href
  {https://ui.adsabs.harvard.edu/abs/2022ApJ...926L..23H} {926, L23}

\bibitem[\protect\citeauthoryear{{Heise}, {Smith}  \& {Calamai}}{{Heise}
  et~al.}{1995}]{HSC95}
{Heise} C.,  {Smith} P.~L.,   {Calamai} A.~G.,  1995, \mn@doi [\apjl]
  {10.1086/309676}, \href
  {https://ui.adsabs.harvard.edu/abs/1995ApJ...451L..41H} {451, L41}

\bibitem[\protect\citeauthoryear{{Henney} \& {Arthur}}{{Henney} \&
  {Arthur}}{1998}]{Henney:1998b}
{Henney} W.~J.,  {Arthur} S.~J.,  1998, \mn@doi [\aj] {10.1086/300433}, \href
  {http://adsabs.harvard.edu/abs/1998AJ....116..322H} {116, 322}

\bibitem[\protect\citeauthoryear{{Henney} \& {O'Dell}}{{Henney} \&
  {O'Dell}}{1999}]{henney99}
{Henney} W.~J.,  {O'Dell} C.~R.,  1999, \mn@doi [\aj] {10.1086/301087}, \href
  {https://ui.adsabs.harvard.edu/abs/1999AJ....118.2350H} {118, 2350}

\bibitem[\protect\citeauthoryear{{Henney}, {O'Dell}, {Meaburn}, {Garrington}
  \& {López}}{{Henney} et~al.}{2002}]{Henney:2002b}
{Henney} W.~J.,  {O'Dell} C.~R.,  {Meaburn} J.,  {Garrington} S.~T.,   {López}
  J.~A.,  2002, \mn@doi [\apj] {10.1086/338055}, \href
  {http://adsabs.harvard.edu/abs/2002ApJ...566..315H} {566, 315}

\bibitem[\protect\citeauthoryear{{Henney}, {O'Dell}, {Zapata},
  {Garc{\'{\i}}a-D{\'{\i}}az}, {Rodr{\'{\i}}guez}  \& {Robberto}}{{Henney}
  et~al.}{2007}]{Henney:2007b}
{Henney} W.~J.,  {O'Dell} C.~R.,  {Zapata} L.~A.,  {Garc{\'{\i}}a-D{\'{\i}}az}
  M.~T.,  {Rodr{\'{\i}}guez} L.~F.,   {Robberto} M.,  2007, \mn@doi [\aj]
  {10.1086/513074}, \href {http://adsabs.harvard.edu/abs/2007AJ....133.2192H}
  {133, 2192}

\bibitem[\protect\citeauthoryear{{Hester} et~al.,}{{Hester}
  et~al.}{1996}]{Hester:1996w}
{Hester} J.~J.,  et~al., 1996, \mn@doi [\aj] {10.1086/117968}, \href
  {https://ui.adsabs.harvard.edu/abs/1996AJ....111.2349H} {111, 2349}

\bibitem[\protect\citeauthoryear{{Hillenbrand}}{{Hillenbrand}}{1997}]{Hillenbrand:1997a}
{Hillenbrand} L.~A.,  1997, \mn@doi [\aj] {10.1086/118389}, \href
  {http://adsabs.harvard.edu/abs/1997AJ....113.1733H} {113, 1733}

\bibitem[\protect\citeauthoryear{{Hudson}, {Ramsbottom}  \& {Scott}}{{Hudson}
  et~al.}{2012a}]{Hudson12}
{Hudson} C.~E.,  {Ramsbottom} C.~A.,   {Scott} M.~P.,  2012a, \mn@doi [\apj]
  {10.1088/0004-637X/750/1/65}, \href
  {https://ui.adsabs.harvard.edu/abs/2012ApJ...750...65H} {750, 65}

\bibitem[\protect\citeauthoryear{{Hudson}, {Ramsbottom}  \& {Scott}}{{Hudson}
  et~al.}{2012b}]{HRS12}
{Hudson} C.~E.,  {Ramsbottom} C.~A.,   {Scott} M.~P.,  2012b, \mn@doi [\apj]
  {10.1088/0004-637X/750/1/65}, \href
  {https://ui.adsabs.harvard.edu/abs/2012ApJ...750...65H} {750, 65}

\bibitem[\protect\citeauthoryear{{Jenkins}}{{Jenkins}}{2009}]{Jenkins09}
{Jenkins} E.~B.,  2009, \mn@doi [\apj] {10.1088/0004-637X/700/2/1299}, \href
  {https://ui.adsabs.harvard.edu/abs/2009ApJ...700.1299J} {700, 1299}

\bibitem[\protect\citeauthoryear{{Johnstone}, {Hollenbach}  \&
  {Bally}}{{Johnstone} et~al.}{1998}]{Johnstone:1998a}
{Johnstone} D.,  {Hollenbach} D.,   {Bally} J.,  1998, \mn@doi [\apj]
  {10.1086/305658}, \href
  {https://ui.adsabs.harvard.edu/abs/1998ApJ...499..758J} {499, 758}

\bibitem[\protect\citeauthoryear{{Juan de Dios} \& {Rodr{\'\i}guez}}{{Juan de
  Dios} \& {Rodr{\'\i}guez}}{2017}]{Juan-de-Dios17}
{Juan de Dios} L.,  {Rodr{\'\i}guez} M.,  2017, \mn@doi [\mnras]
  {10.1093/mnras/stx916}, \href
  {https://ui.adsabs.harvard.edu/abs/2017MNRAS.469.1036J} {469, 1036}

\bibitem[\protect\citeauthoryear{{Kama}, {Shorttle}, {Jermyn}, {Folsom},
  {Furuya}, {Bergin}, {Walsh}  \& {Keller}}{{Kama} et~al.}{2019}]{Kama19}
{Kama} M.,  {Shorttle} O.,  {Jermyn} A.~S.,  {Folsom} C.~P.,  {Furuya} K.,
  {Bergin} E.~A.,  {Walsh} C.,   {Keller} L.,  2019, \mn@doi [\apj]
  {10.3847/1538-4357/ab45f8}, \href
  {https://ui.adsabs.harvard.edu/abs/2019ApJ...885..114K} {885, 114}

\bibitem[\protect\citeauthoryear{{Kaufman} \& {Sugar}}{{Kaufman} \&
  {Sugar}}{1986}]{KS86}
{Kaufman} V.,  {Sugar} J.,  1986, \mn@doi [Journal of Physical and Chemical
  Reference Data] {10.1063/1.555775}, \href
  {https://ui.adsabs.harvard.edu/abs/1986JPCRD..15..321K} {15, 321}

\bibitem[\protect\citeauthoryear{{Kim}, {Lu}, {Konopacky}, {Chu}, {Toller},
  {Anderson}, {Theissen}  \& {Morris}}{{Kim} et~al.}{2019}]{Kim:2019a}
{Kim} D.,  {Lu} J.~R.,  {Konopacky} Q.,  {Chu} L.,  {Toller} E.,  {Anderson}
  J.,  {Theissen} C.~A.,   {Morris} M.~R.,  2019, \mn@doi [\aj]
  {10.3847/1538-3881/aafb09}, \href
  {https://ui.adsabs.harvard.edu/abs/2019AJ....157..109K} {157, 109}

\bibitem[\protect\citeauthoryear{{Kunth} \& {Sargent}}{{Kunth} \&
  {Sargent}}{1983}]{kunthsargent83}
{Kunth} D.,  {Sargent} W.~L.~W.,  1983, \mn@doi [\apj] {10.1086/161350}, \href
  {https://ui.adsabs.harvard.edu/abs/1983ApJ...273...81K} {273, 81}

\bibitem[\protect\citeauthoryear{{LaJohn} \& {Luke}}{{LaJohn} \&
  {Luke}}{1993}]{LL93}
{LaJohn} L.,  {Luke} T.~M.,  1993, \mn@doi [\physscr]
  {10.1088/0031-8949/47/4/011}, \href
  {https://ui.adsabs.harvard.edu/abs/1993PhyS...47..542L} {47, 542}

\bibitem[\protect\citeauthoryear{{Laques} \& {Vidal}}{{Laques} \&
  {Vidal}}{1979}]{Laques:1979a}
{Laques} P.,  {Vidal} J.~L.,  1979, \aap, \href
  {http://adsabs.harvard.edu/abs/1979A%26A....73...97L} {73, 97}

\bibitem[\protect\citeauthoryear{{Lee}, {Ho}, {Li}, {Hirano}, {Zhang}  \&
  {Shang}}{{Lee} et~al.}{2017}]{Lee:2017r}
{Lee} C.-F.,  {Ho} P. T.~P.,  {Li} Z.-Y.,  {Hirano} N.,  {Zhang} Q.,   {Shang}
  H.,  2017, \mn@doi [Nature Astronomy] {10.1038/s41550-017-0152}, \href
  {https://ui.adsabs.harvard.edu/abs/2017NatAs...1E.152L} {1, 0152}

\bibitem[\protect\citeauthoryear{{Lodders}}{{Lodders}}{2019}]{lodders19}
{Lodders} K.,  2019, arXiv e-prints, \href
  {https://ui.adsabs.harvard.edu/abs/2019arXiv191200844L} {p. arXiv:1912.00844}

\bibitem[\protect\citeauthoryear{{Luridiana}, {Morisset}  \&
  {Shaw}}{{Luridiana} et~al.}{2015}]{Luridiana15}
{Luridiana} V.,  {Morisset} C.,   {Shaw} R.~A.,  2015, \mn@doi [\aap]
  {10.1051/0004-6361/201323152}, \href
  {https://ui.adsabs.harvard.edu/abs/2015A&A...573A..42L} {573, A42}

\bibitem[\protect\citeauthoryear{{McBride} \& {Kounkel}}{{McBride} \&
  {Kounkel}}{2019}]{McBride:2019a}
{McBride} A.,  {Kounkel} M.,  2019, \mn@doi [\apj] {10.3847/1538-4357/ab3df9},
  \href {https://ui.adsabs.harvard.edu/abs/2019ApJ...884....6M} {884, 6}

\bibitem[\protect\citeauthoryear{{M{\'e}ndez-Delgado}, {Esteban},
  {Garc{\'\i}a-Rojas}, {Henney}, {Mesa-Delgado}  \&
  {Arellano-C{\'o}rdova}}{{M{\'e}ndez-Delgado} et~al.}{2021a}]{mendez2021}
{M{\'e}ndez-Delgado} J.~E.,  {Esteban} C.,  {Garc{\'\i}a-Rojas} J.,  {Henney}
  W.~J.,  {Mesa-Delgado} A.,   {Arellano-C{\'o}rdova} K.~Z.,  2021a, \mn@doi
  [\mnras] {10.1093/mnras/stab068}, \href
  {https://ui.adsabs.harvard.edu/abs/2021MNRAS.502.1703M} {502, 1703}

\bibitem[\protect\citeauthoryear{{M{\'e}ndez-Delgado}, {Henney}, {Esteban},
  {Garc{\'\i}a-Rojas}, {Mesa-Delgado}  \&
  {Arellano-C{\'o}rdova}}{{M{\'e}ndez-Delgado} et~al.}{2021b}]{mendez2021-2}
{M{\'e}ndez-Delgado} J.~E.,  {Henney} W.~J.,  {Esteban} C.,
  {Garc{\'\i}a-Rojas} J.,  {Mesa-Delgado} A.,   {Arellano-C{\'o}rdova} K.~Z.,
  2021b, \mn@doi [\apj] {10.3847/1538-4357/ac0cf5}, \href
  {https://ui.adsabs.harvard.edu/abs/2021ApJ...918...27M} {918, 27}

\bibitem[\protect\citeauthoryear{{Mendoza} \& {Zeippen}}{{Mendoza} \&
  {Zeippen}}{1982}]{MZ82b}
{Mendoza} C.,  {Zeippen} C.~J.,  1982, \mn@doi [\mnras]
  {10.1093/mnras/199.4.1025}, \href
  {https://ui.adsabs.harvard.edu/abs/1982MNRAS.199.1025M} {199, 1025}

\bibitem[\protect\citeauthoryear{{Mesa-Delgado}, {Esteban},
  {Garc{\'\i}a-Rojas}, {Luridiana}, {Bautista}, {Rodr{\'\i}guez},
  {L{\'o}pez-Mart{\'\i}n}  \& {Peimbert}}{{Mesa-Delgado}
  et~al.}{2009}]{mesadelgado09}
{Mesa-Delgado} A.,  {Esteban} C.,  {Garc{\'\i}a-Rojas} J.,  {Luridiana} V.,
  {Bautista} M.,  {Rodr{\'\i}guez} M.,  {L{\'o}pez-Mart{\'\i}n} L.,
  {Peimbert} M.,  2009, \mn@doi [\mnras] {10.1111/j.1365-2966.2009.14554.x},
  \href {https://ui.adsabs.harvard.edu/abs/2009MNRAS.395..855M} {395, 855}

\bibitem[\protect\citeauthoryear{{Moehler}, {Dreizler}, {LeBlanc}, {Khalack},
  {Michaud}, {Richer}, {Sweigart}  \& {Grundahl}}{{Moehler}
  et~al.}{2014a}]{Moehler14a}
{Moehler} S.,  {Dreizler} S.,  {LeBlanc} F.,  {Khalack} V.,  {Michaud} G.,
  {Richer} J.,  {Sweigart} A.~V.,   {Grundahl} F.,  2014a, \mn@doi [\aap]
  {10.1051/0004-6361/201322953}, \href
  {https://ui.adsabs.harvard.edu/abs/2014A&A...565A.100M} {565, A100}

\bibitem[\protect\citeauthoryear{{Moehler} et~al.,}{{Moehler}
  et~al.}{2014b}]{Moehler14b}
{Moehler} S.,  et~al., 2014b, \mn@doi [\aap] {10.1051/0004-6361/201423790},
  \href {https://ui.adsabs.harvard.edu/abs/2014A&A...568A...9M} {568, A9}

\bibitem[\protect\citeauthoryear{{Nisini}, {Caratti o Garatti}, {Giannini}  \&
  {Lorenzetti}}{{Nisini} et~al.}{2002}]{Nisini:2002f}
{Nisini} B.,  {Caratti o Garatti} A.,  {Giannini} T.,   {Lorenzetti} D.,  2002,
  \mn@doi [\aap] {10.1051/0004-6361:20021062}, \href
  {https://ui.adsabs.harvard.edu/abs/2002A&A...393.1035N} {393, 1035}

\bibitem[\protect\citeauthoryear{{Noll}, {Kausch}, {Barden}, {Jones},
  {Szyszka}, {Kimeswenger}  \& {Vinther}}{{Noll} et~al.}{2012}]{Noll12}
{Noll} S.,  {Kausch} W.,  {Barden} M.,  {Jones} A.~M.,  {Szyszka} C.,
  {Kimeswenger} S.,   {Vinther} J.,  2012, \mn@doi [\aap]
  {10.1051/0004-6361/201219040}, \href
  {https://ui.adsabs.harvard.edu/abs/2012A&A...543A..92N} {543, A92}

\bibitem[\protect\citeauthoryear{{O'Dell} \& {Henney}}{{O'Dell} \&
  {Henney}}{2008}]{odellyhenney08}
{O'Dell} C.~R.,  {Henney} W.~J.,  2008, \mn@doi [\aj]
  {10.1088/0004-6256/136/4/1566}, \href
  {https://ui.adsabs.harvard.edu/abs/2008AJ....136.1566O} {136, 1566}

\bibitem[\protect\citeauthoryear{{O'Dell} \& {Wen}}{{O'Dell} \&
  {Wen}}{1994}]{Odell:1994a}
{O'Dell} C.~R.,  {Wen} Z.,  1994, \mn@doi [\apj] {10.1086/174892}, \href
  {https://ui.adsabs.harvard.edu/abs/1994ApJ...436..194O} {436, 194}

\bibitem[\protect\citeauthoryear{{O'Dell}, {Wen}  \& {Hu}}{{O'Dell}
  et~al.}{1993}]{Odell1993}
{O'Dell} C.~R.,  {Wen} Z.,   {Hu} X.,  1993, \mn@doi [\apj] {10.1086/172786},
  \href {https://ui.adsabs.harvard.edu/abs/1993ApJ...410..696O} {410, 696}

\bibitem[\protect\citeauthoryear{{O'Dell}, {Ferland}, {Henney}, {Peimbert},
  {Garc{\'\i}a-D{\'\i}az}  \& {Rubin}}{{O'Dell} et~al.}{2015}]{Odell15}
{O'Dell} C.~R.,  {Ferland} G.~J.,  {Henney} W.~J.,  {Peimbert} M.,
  {Garc{\'\i}a-D{\'\i}az} M.~T.,   {Rubin} R.~H.,  2015, \mn@doi [\aj]
  {10.1088/0004-6256/150/4/108}, \href
  {https://ui.adsabs.harvard.edu/abs/2015AJ....150..108O} {150, 108}

\bibitem[\protect\citeauthoryear{{Pascual}, {Cardiel}, {Gil de Paz}, {Carasco},
  {Gallego}, {Iglesias-P{\'a}ramo}  \& {Cedazo}}{{Pascual}
  et~al.}{2019}]{pascual+19}
{Pascual} S.,  {Cardiel} N.,  {Gil de Paz} A.,  {Carasco} E.,  {Gallego} J.,
  {Iglesias-P{\'a}ramo} J.,   {Cedazo} R.,  2019, in {Montesinos} B.,  {Asensio
  Ramos} A.,  {Buitrago} F.,  {Sch{\"o}del} R.,  {Villaver} E.,
  {P{\'e}rez-Hoyos} S.,   {Ord{\'o}{\~n}ez-Etxeberria} I.,  eds, Highlights on
  Spanish Astrophysics X. pp 227--227

\bibitem[\protect\citeauthoryear{{Pascual}, {Cardiel}, {Picazo-Sanchez},
  {Castillo-Morales}  \& {Gil De Paz}}{{Pascual} et~al.}{2021}]{pascual+20}
{Pascual} S.,  {Cardiel} N.,  {Picazo-Sanchez} P.,  {Castillo-Morales} A.,
  {Gil De Paz} A.,  2021, {guaix-ucm/megaradrp: v0.11},
  \mn@doi{10.5281/zenodo.593647}

\bibitem[\protect\citeauthoryear{{Pinilla}, {Benisty}  \&
  {Birnstiel}}{{Pinilla} et~al.}{2012}]{Pinilla2012}
{Pinilla} P.,  {Benisty} M.,   {Birnstiel} T.,  2012, \mn@doi [\aap]
  {10.1051/0004-6361/201219315}, \href
  {https://ui.adsabs.harvard.edu/abs/2012A&A...545A..81P} {545, A81}

\bibitem[\protect\citeauthoryear{{Platais} et~al.,}{{Platais}
  et~al.}{2020}]{Platais:2020f}
{Platais} I.,  et~al., 2020, \mn@doi [\aj] {10.3847/1538-3881/ab8d42}, \href
  {https://ui.adsabs.harvard.edu/abs/2020AJ....159..272P} {159, 272}

\bibitem[\protect\citeauthoryear{{Podio}, {Medves}, {Bacciotti}, {Eislöffel}
  \& {Ray}}{{Podio} et~al.}{2009}]{Podio:2009a}
{Podio} L.,  {Medves} S.,  {Bacciotti} F.,  {Eislöffel} J.,   {Ray} T.,  2009,
  \mn@doi [\aap] {10.1051/0004-6361/200912408}, \href
  {https://ui.adsabs.harvard.edu/abs/2009A&A...506..779P} {506, 779}

\bibitem[\protect\citeauthoryear{{Podobedova}, {Kelleher}  \&
  {Wiese}}{{Podobedova} et~al.}{2009}]{Podobedova09}
{Podobedova} L.~I.,  {Kelleher} D.~E.,   {Wiese} W.~L.,  2009, \mn@doi [Journal
  of Physical and Chemical Reference Data] {10.1063/1.3032939}, \href
  {https://ui.adsabs.harvard.edu/abs/2009JPCRD..38..171P} {38, 171}

\bibitem[\protect\citeauthoryear{{Robberto} et~al.,}{{Robberto}
  et~al.}{2013}]{Robberto:2013a}
{Robberto} M.,  et~al., 2013, \mn@doi [\apjs] {10.1088/0067-0049/207/1/10},
  \href {https://ui.adsabs.harvard.edu/abs/2013ApJS..207...10R} {207, 10}

\bibitem[\protect\citeauthoryear{{Rodenkirch} \& {Dullemond}}{{Rodenkirch} \&
  {Dullemond}}{2022}]{Rodenkirch:2022o}
{Rodenkirch} P.~J.,  {Dullemond} C.~P.,  2022, \mn@doi [\aap]
  {10.1051/0004-6361/202142571}, \href
  {https://ui.adsabs.harvard.edu/abs/2022A&A...659A..42R} {659, A42}

\bibitem[\protect\citeauthoryear{{Rodr{\'{\i}}guez}}{{Rodr{\'{\i}}guez}}{2002}]{Rodriguez:2002a}
{Rodr{\'{\i}}guez} M.,  2002, \mn@doi [\aap] {10.1051/0004-6361:20011823},
  \href {http://adsabs.harvard.edu/abs/2002A%26A...389..556R} {389, 556}

\bibitem[\protect\citeauthoryear{{Rodr{\'\i}guez}}{{Rodr{\'\i}guez}}{2020}]{rodriguez20}
{Rodr{\'\i}guez} M.,  2020, \mn@doi [\mnras] {10.1093/mnras/staa1286}, \href
  {https://ui.adsabs.harvard.edu/abs/2020MNRAS.495.1016R} {495, 1016}

\bibitem[\protect\citeauthoryear{{Rodr{\'\i}guez} \& {Rubin}}{{Rodr{\'\i}guez}
  \& {Rubin}}{2005}]{rodriguez05}
{Rodr{\'\i}guez} M.,  {Rubin} R.~H.,  2005, \mn@doi [\apj] {10.1086/429958},
  \href {https://ui.adsabs.harvard.edu/abs/2005ApJ...626..900R} {626, 900}

\bibitem[\protect\citeauthoryear{{Smith}, {Bally}, {Shuping}, {Morris}  \&
  {Kassis}}{{Smith} et~al.}{2005}]{smith05}
{Smith} N.,  {Bally} J.,  {Shuping} R.~Y.,  {Morris} M.,   {Kassis} M.,  2005,
  \mn@doi [\aj] {10.1086/432912}, \href
  {https://ui.adsabs.harvard.edu/abs/2005AJ....130.1763S} {130, 1763}

\bibitem[\protect\citeauthoryear{{Sota}, {Ma{\'{\i}}z Apellániz}, {Walborn},
  {Alfaro}, {Barbá}, {Morrell}, {Gamen}  \& {Arias}}{{Sota}
  et~al.}{2011}]{Sota:2011a}
{Sota} A.,  {Ma{\'{\i}}z Apellániz} J.,  {Walborn} N.~R.,  {Alfaro} E.~J.,
  {Barbá} R.~H.,  {Morrell} N.~I.,  {Gamen} R.~C.,   {Arias} J.~I.,  2011,
  \mn@doi [\apjs] {10.1088/0067-0049/193/2/24}, \href
  {http://adsabs.harvard.edu/abs/2011ApJS..193...24S} {193, 24}

\bibitem[\protect\citeauthoryear{{Tayal}}{{Tayal}}{1997}]{Tayal97}
{Tayal} S.~S.,  1997, \mn@doi [Atomic Data and Nuclear Data Tables]
  {10.1006/adnd.1997.0753}, \href
  {https://ui.adsabs.harvard.edu/abs/1997ADNDT..67..331T} {67, 331}

\bibitem[\protect\citeauthoryear{{Tayal} \& {Gupta}}{{Tayal} \&
  {Gupta}}{1999}]{TG99}
{Tayal} S.~S.,  {Gupta} G.~P.,  1999, \mn@doi [\apj] {10.1086/307971}, \href
  {https://ui.adsabs.harvard.edu/abs/1999ApJ...526..544T} {526, 544}

\bibitem[\protect\citeauthoryear{{Tayal}, {Zatsarinny}  \& {Sossah}}{{Tayal}
  et~al.}{2019}]{TZS19}
{Tayal} S.~S.,  {Zatsarinny} O.,   {Sossah} A.~M.,  2019, \mn@doi [\apjs]
  {10.3847/1538-4365/ab17e1}, \href
  {https://ui.adsabs.harvard.edu/abs/2019ApJS..242....9T} {242, 9}

\bibitem[\protect\citeauthoryear{{Theissen}, {Konopacky}, {Lu}, {Kim}, {Zhang},
  {Hsu}, {Chu}  \& {Wei}}{{Theissen} et~al.}{2022}]{Theissen:2022k}
{Theissen} C.~A.,  {Konopacky} Q.~M.,  {Lu} J.~R.,  {Kim} D.,  {Zhang} S.~Y.,
  {Hsu} C.-C.,  {Chu} L.,   {Wei} L.,  2022, \mn@doi [\apj]
  {10.3847/1538-4357/ac3252}, \href
  {https://ui.adsabs.harvard.edu/abs/2022ApJ...926..141T} {926, 141}

\bibitem[\protect\citeauthoryear{{Tobin}, {Hartmann}, {Furesz}, {Mateo}  \&
  {Megeath}}{{Tobin} et~al.}{2009}]{Tobin:2009a}
{Tobin} J.~J.,  {Hartmann} L.,  {Furesz} G.,  {Mateo} M.,   {Megeath} S.~T.,
  2009, \mn@doi [\apj] {10.1088/0004-637X/697/2/1103}, \href
  {http://adsabs.harvard.edu/abs/2009ApJ...697.1103T} {697, 1103}

\bibitem[\protect\citeauthoryear{{Tody}}{{Tody}}{1993}]{Tody93}
{Tody} D.,  1993, {IRAF in the Nineties}.
p.~173

\bibitem[\protect\citeauthoryear{{Vicente}, {Bern{\'e}}, {Tielens},
  {Hu{\'e}lamo}, {Pantin}, {Kamp}  \& {Carmona}}{{Vicente}
  et~al.}{2013}]{Vicente:2013a}
{Vicente} S.,  {Bern{\'e}} O.,  {Tielens} A.~G.~G.~M.,  {Hu{\'e}lamo} N.,
  {Pantin} E.,  {Kamp} I.,   {Carmona} A.,  2013, \mn@doi [\apjl]
  {10.1088/2041-8205/765/2/L38}, \href
  {https://ui.adsabs.harvard.edu/abs/2013ApJ...765L..38V} {765, L38}

\bibitem[\protect\citeauthoryear{{Weilbacher} et~al.,}{{Weilbacher}
  et~al.}{2015}]{Weilbacher15}
{Weilbacher} P.~M.,  et~al., 2015, \mn@doi [\aap]
  {10.1051/0004-6361/201526529}, \href
  {https://ui.adsabs.harvard.edu/abs/2015A&A...582A.114W} {582, A114}

\bibitem[\protect\citeauthoryear{{Williams}, {Andrews}  \& {Wilner}}{{Williams}
  et~al.}{2005}]{Williams:2005a}
{Williams} J.~P.,  {Andrews} S.~M.,   {Wilner} D.~J.,  2005, \mn@doi [\apj]
  {10.1086/444493}, \href
  {https://ui.adsabs.harvard.edu/abs/2005ApJ...634..495W} {634, 495}

\makeatother
\end{thebibliography}

\newpage


\appendix

\section{Some extra material}
\label{sec:apendix_A}

\begin{table*}
\caption{Sample of 15 lines of the HH~514~jet base spectrum. All line intensity ratios are referred to F$\left( \mbox{H}\beta \right)$ or I$\left( \mbox{H}\beta \right)$ = 100. This is an example of the content found in the supplementary material for all spectra.}
\label{tab:sample_spectra}
\begin{tabular}{ccccccccccccccccccccc}
\hline
$\lambda_0$ ( \AA ) & Ion & $\lambda_{\text{obs}}$( \AA ) & $v_r$ (Km s$^{-1}$) & FWHM (Km s$^{-1}$) & F$\left( \lambda \right)$/F$\left( \mbox{H}\beta \right)$ & I$\left( \lambda \right)$/I$\left( \mbox{H}\beta \right)$ & Err \% & Notes \\
\hline
4363.21 & [O\thinspace III] & 4365.44 & 153 & 19.7 $\pm$ 1.4 & 2.10 & 2.48 & 6 & \\
4471.47 & He\thinspace I & 4473.78 & 155 & 22.4 $\pm$ 1.1 & 4.00 & 4.54 & 4 & \\
4596.84 & [Ni\thinspace III] & 4599.38 & 166 & 10 $\pm$ 6 & 0.24 & 0.26 & 38 & \\
4607.12 & [Fe\thinspace III] & 4609.42 & 150 & 19.8 $\pm$ 2.0 & 1.50 & 1.62 & 8 & \\
4658.17 & [Fe\thinspace III] & 4660.51 & 151 & 18.85 $\pm$ 0.07 & 16.85 & 18.0 & 2 & \\
4667.11 & [Fe\thinspace III] & 4669.40 & 147 & 24.5 $\pm$ 2.5 & 1.69 & 1.80 & 8 & \\
4701.64 & [Fe\thinspace III] & 4704.00 & 150 & 19.4 $\pm$ 0.4 & 8.27 & 8.69 & 3 & \\
4734.00 & [Fe\thinspace III] & 4736.33 & 148 & 21.5 $\pm$ 1.2 & 4.18 & 4.34 & 4 & \\
4754.81 & [Fe\thinspace III] & 4757.19 & 150 & 19.6 $\pm$ 1.0 & 3.39 & 3.50 & 4 & \\
4769.53 & [Fe\thinspace III] & 4771.92 & 150 & 20.4 $\pm$ 0.8 & 3.36 & 3.45 & 5 & \\
4777.70 & [Fe\thinspace III] & 4780.16 & 154 & 17.6 $\pm$ 1.7 & 2.06 & 2.12 & 7 & \\
4814.54 & [Fe\thinspace II] & 4817.00 & 153 & 17 $\pm$ 6 & 0.50 & 0.51 & 22 & \\
4861.32 & H\thinspace I & 4863.81 & 154 & 27.74 $\pm$ 0.27 & 100.0 & 100.0 & 2 & \\
4881.07 & [Fe\thinspace III] & 4883.58 & 154 & 18.4 $\pm$ 0.8 & 4.18 & 4.15 & 4 & \\
4930.64 & [Fe\thinspace III] & 4933.18 & 154 & 20.8 $\pm$ 3.4 & 1.56 & 1.53 & 12 & blend with [O\thinspace III] 4931.23\\
\hline
\end{tabular}
\end{table*}

\begin{figure*}
  \begin{minipage}{7.5cm}
    \centering\includegraphics[height=4cm,width=\columnwidth]{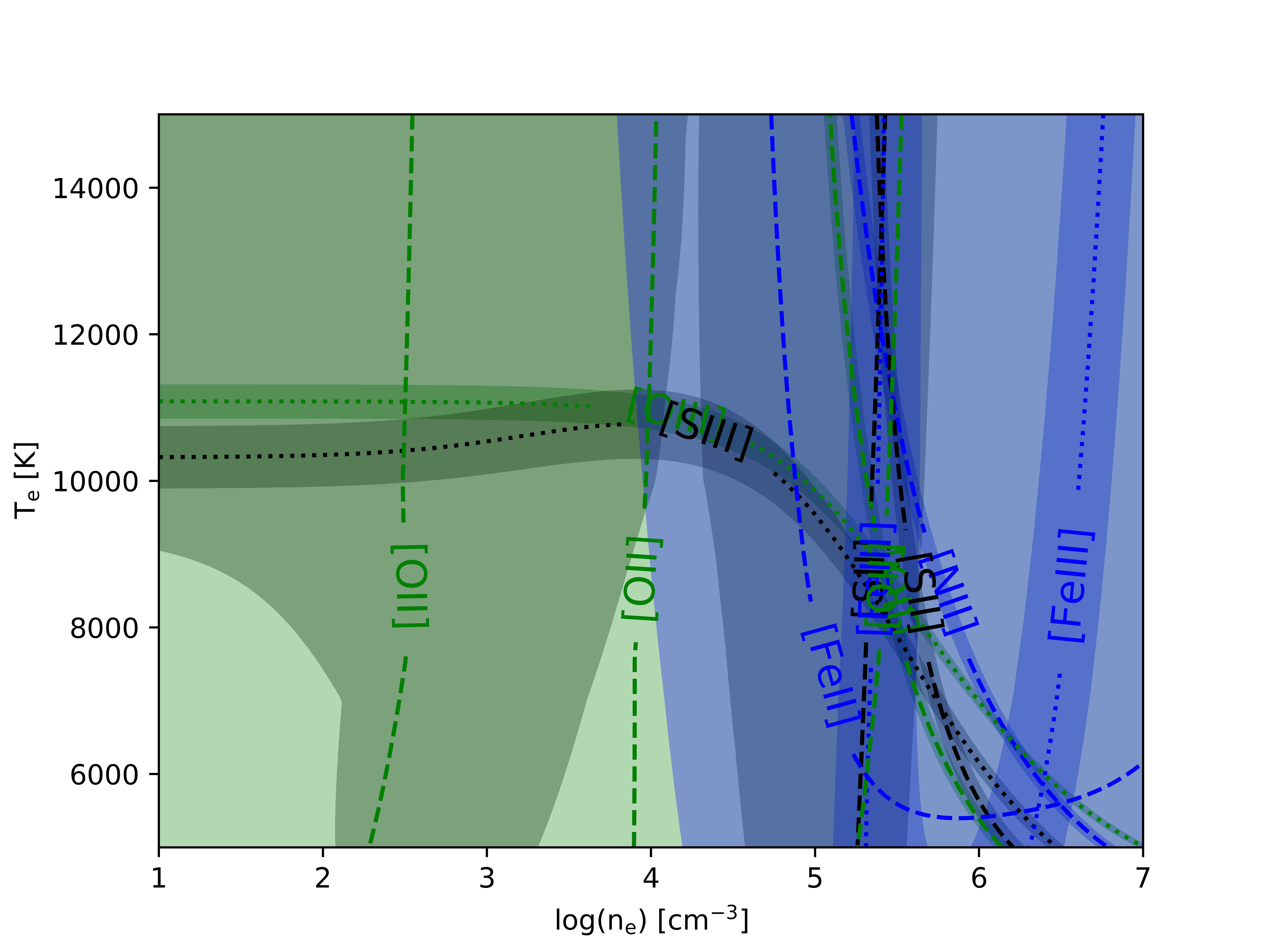}
    \centerline{(a) Cut~1, HH~514 jet base.}
  \end{minipage}
  \begin{minipage}{7.5cm}
     \centering\includegraphics[height=4cm,width=\columnwidth]{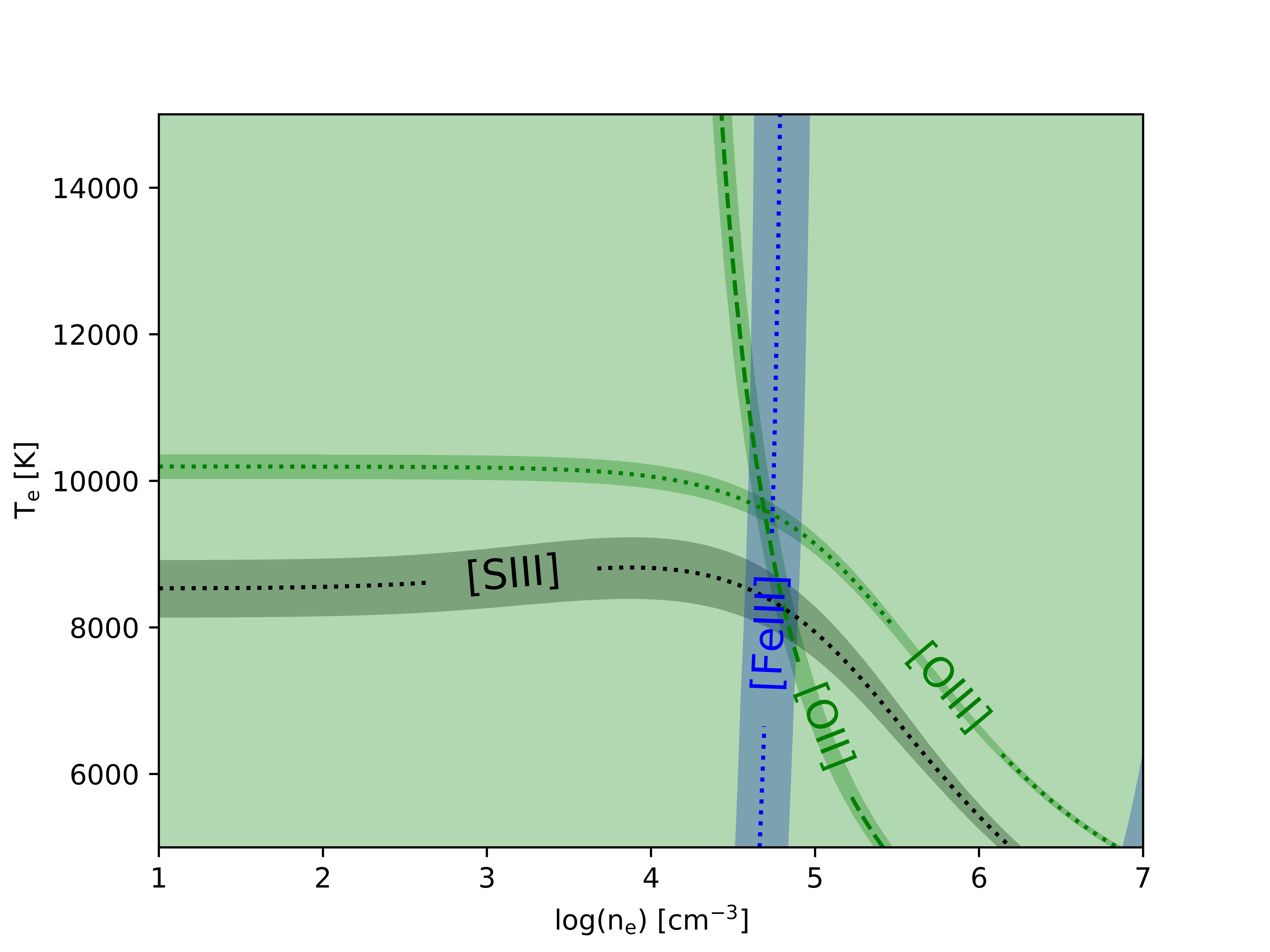}
    \centerline{(b) Cut~2, HH~514 knot.}
  \end{minipage}
 
  \begin{minipage}{7.5cm}
   \centering\includegraphics[height=4cm,width=\columnwidth]{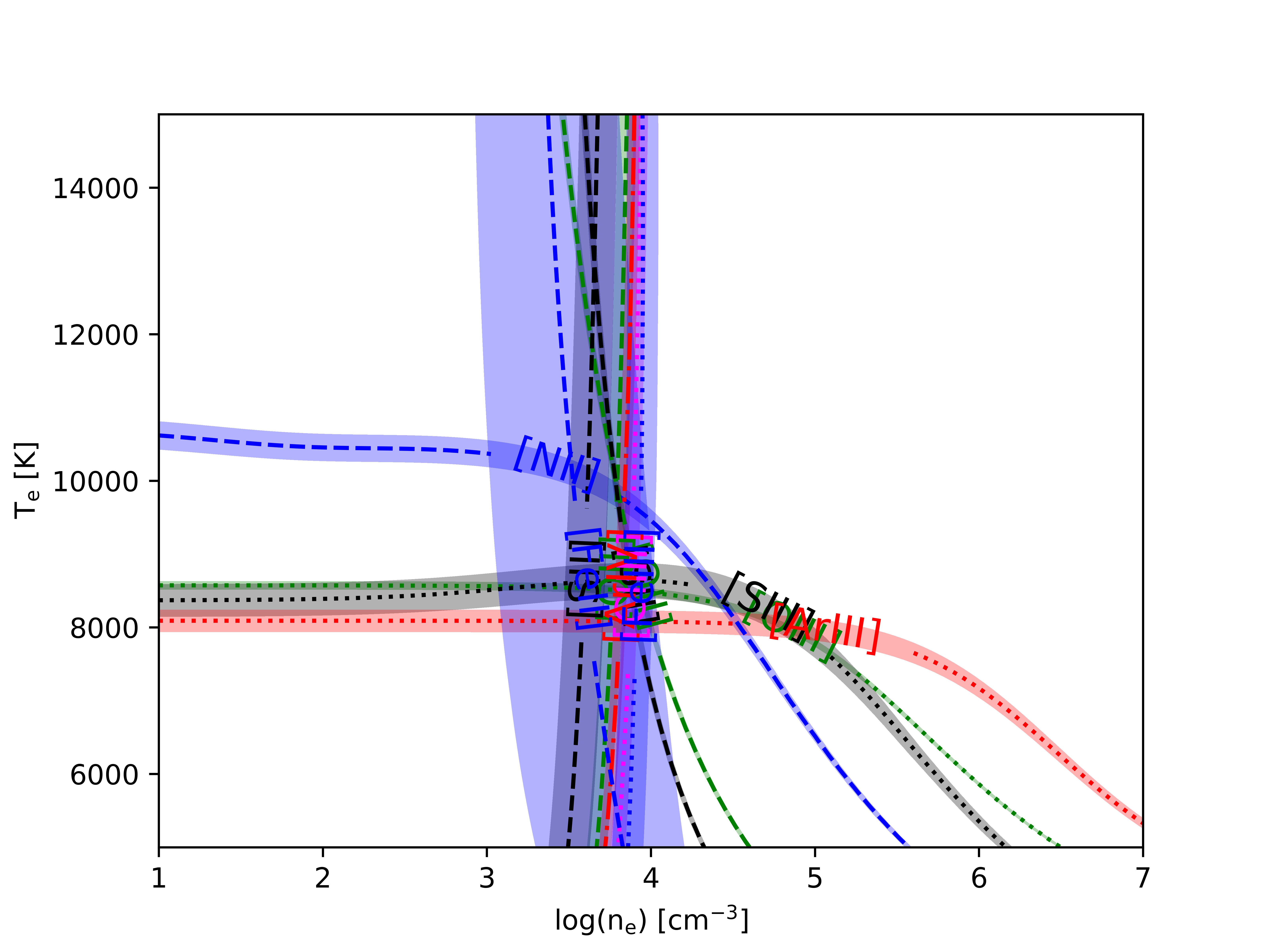}
   \centerline{(c) Cut~2, nebular component.}
  \end{minipage}
  \begin{minipage}{7.5cm}
    \centering\includegraphics[height=4cm,width=\columnwidth]{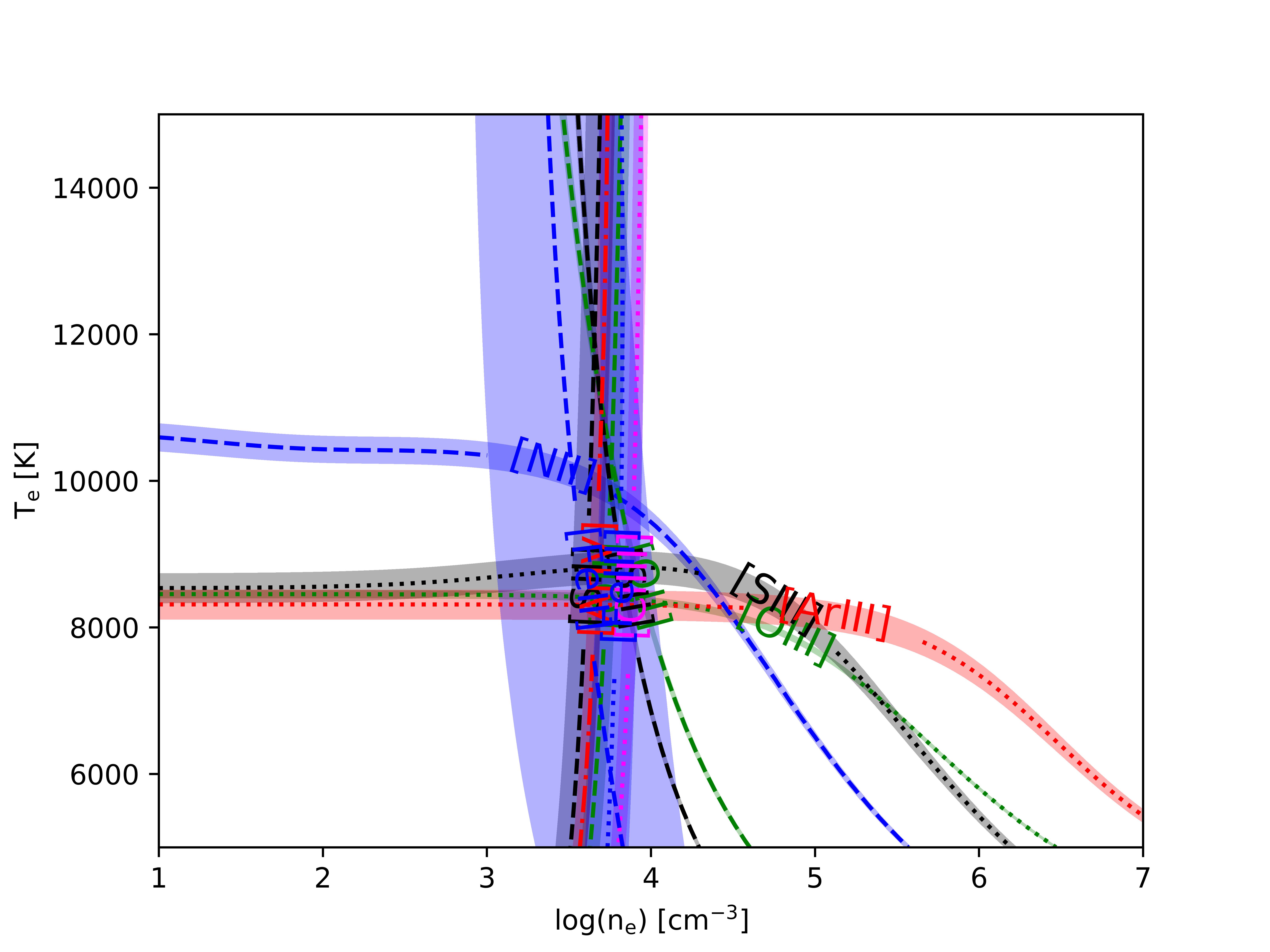}
    \centerline{(d) Cut~3, nebular component.}
  \end{minipage}
  \caption{Plasma diagnostic plots for the individual analyzed components.}
\label{fig:plasma}
\end{figure*}

\begin{figure}
\includegraphics[width=\columnwidth]{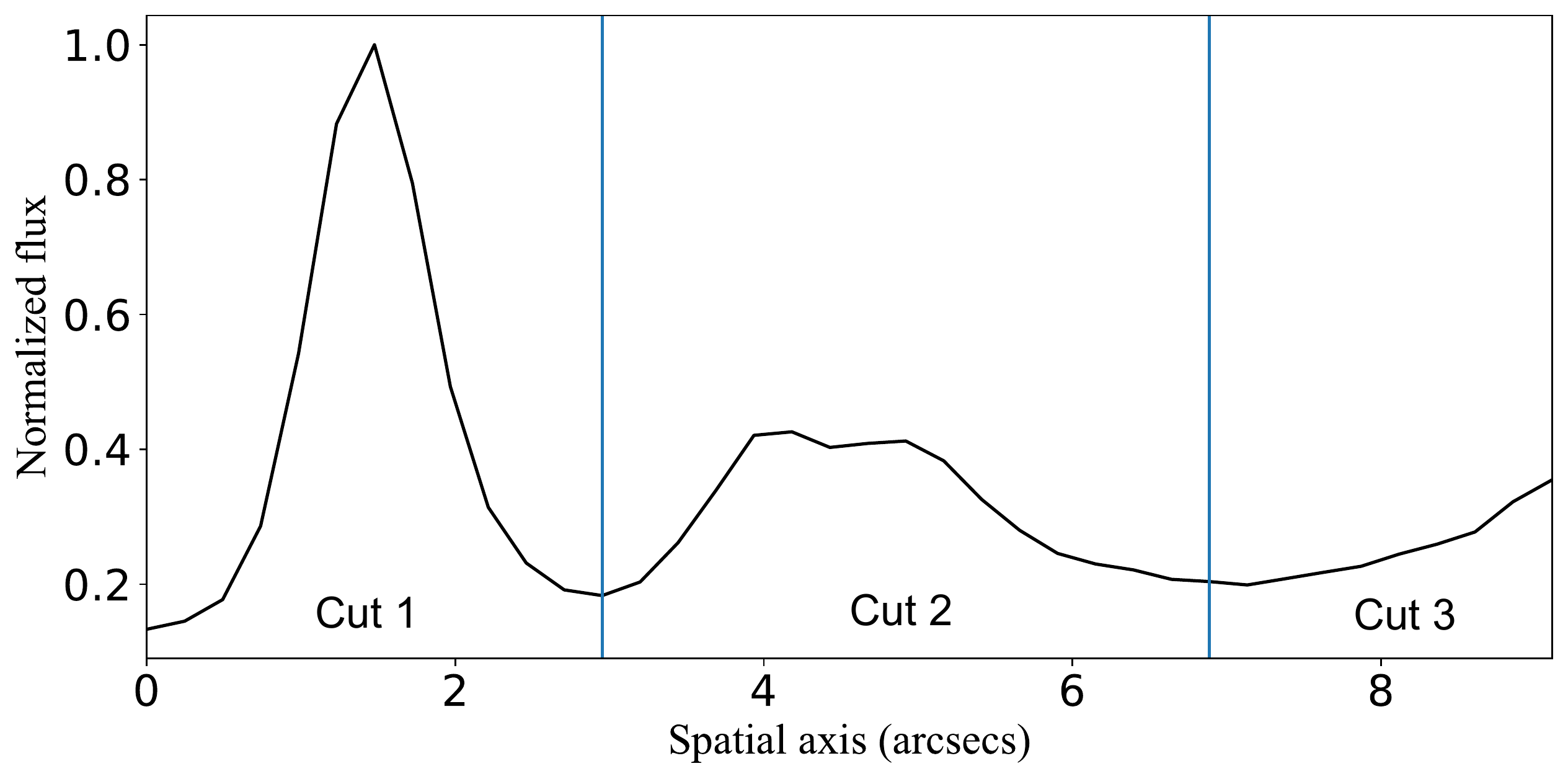}
\caption{Spatial distribution along the slit of the intensity of [Fe\thinspace III] $\lambda 4658$ line centred at a heliocentric velocity of $\sim 150 \text{ km s}^{-1}$.  The emission has been extracted from a window $\sim 45 \text{ km s}^{-1}$ wide }
\label{fig:spatial_dis}
\end{figure}

\begin{table*}
\centering
\caption{[S\thinspace III] line intensity ratios with a common upper level observed in this paper and other objects from the literature and predicted by theoretical models.}
\label{tab:atomic_data_test}
\begin{tabular}{ccccccccccccc}
\hline
Reference & 3722/6312 & 9531/9069 & 9531/8829 &9069/8829 \\
\hline

 & \multicolumn{4}{c}{{\bf Orion Nebula}}\\

\citet{Esteban04} & - & $2.39 \pm 0.51$ & $5600 \pm 1300$ & $2300 \pm 500$\\

\citet{mesadelgado09} & - & $2.32 \pm 0.19$ & - & - \\

\citet{mendez2021} Cut~1 & $0.60 \pm 0.03$ & $2.65 \pm 0.24$ & $8400 \pm 1200$ & $3200 \pm 500$\\

\citet{mendez2021} Cut~2 & - &$2.84 \pm 0.27$ & $7700 \pm 900$& $2730 \pm 310$\\

\citet{mendez2021} Cut~3 & - &$2.88 \pm 0.26$&$8300 \pm 1100$&$2900 \pm 400$\\

\citet{mendez2021} Cut~4 & - &$2.50 \pm 0.24$&$7700 \pm 1300$ & $3100 \pm 500$ \\

\citet{mendez2021-2} Cut~1 + NIL & - & $2.43 \pm 0.18$ &$5800 \pm 700$&$2400\pm 270$ \\

\citet{mendez2021-2} Cut~2 & - & $2.44 \pm 0.22$ & $6800\pm 1200$& $2800 \pm 500$ \\

\citet{mendez2021-2} NIL & - &$2.65 \pm 0.46$ &-&- \\

This work Cut~2 & $0.64 \pm 0.02$ & $2.69 \pm 0.28$ & $7800 \pm 1100$ & $2900 \pm 400$\\

This work Cut~3 & - & $2.60 \pm 0.27$ & $7400 \pm 1000$ & $2800 \pm 400$\\

& \multicolumn{4}{c}{{\bf Photoionized Herbig-Haro Objects}}\\

\citet{mesadelgado09} HH~202~S & - & $2.54 \pm 0.22$ & -&-\\

\citet{mendez2021} HH~529~II &-&$2.46 \pm 0.22$&$7500 \pm 1800$&$3100 \pm 700$\\

\citet{mendez2021} HH~529~III &-&$2.45 \pm 0.29$ & - & -\\

\citet{mendez2021-2} HH~204 & - & $2.39 \pm 0.14$ & $8700 \pm 1000$ & $3600 \pm 400$\\

This work HH~514~jet base & - & $2.60 \pm 0.32$ &-&-\\

This work HH~514~knot & - & $2.55 \pm 0.38$ &-&-\\

 & \multicolumn{4}{c}{{\bf Galactic H~II regions}}\\

\citet{garciarojas04} NGC~3576 & -& - & - & $2500 \pm 400$ \\

\citet{garciarojas05} Sh~2-311& - & $2.70\pm 0.21$&-&-\\

\citet{garciarojas06} M~16 &-&$ 2.43\pm 0.29$&-&-\\

\citet{garciarojas06} M~20 & -& $ 2.29\pm 0.20$&-&-\\

\citet{garciarojas06} NGC~3603 & - & - & $5900 \pm 800$ & - \\

\citet{garciarojas07-2} M~8 &-& - &- & $2700 \pm 700$  \\

\citet{garciarojas07-2} M~17 &-& $2.51 \pm 0.25$ &-&-\\

\citet{Esteban13} NGC~2579 &- &$2.18 \pm 0.12$ &-&-\\


{\bf Adopted observed value} & \boldmath $0.62 \pm 0.03$ & \boldmath $2.45 \pm 0.18$& \boldmath$ 7200\pm 1200$ & \boldmath $2900\pm400$\\

 & \multicolumn{4}{c}{{\bf Theoretical Predictions}}\\

LL93-HSC95-MZ82b-KS86 &0.52& 5.52&14963&2710\\

MZ82b-HSC95-LL93  & 0.61 & 2.48 &9169&3697\\

FFTI06 &0.54& 2.47&8471&3431\\

TZS19 & 0.50 & 2.54&5505&2171\\

CHIANTI & 0.53 & 2.51&8474&3373\\

\hline
\end{tabular}
\end{table*}

\begin{table*}
\centering
\caption{[S\thinspace III] $\lambda \lambda 9069, 9531$ line intensities with respect to H$\beta$ = 100.0 observed in this paper and other objects from the literature corrected from telluric absorption bands.}
\label{tab:tellabs}
\begin{tabular}{ccccccccccccc}
\hline
Object &   \multicolumn{2}{c}{$\lambda 9069$} & \multicolumn{2}{c}{$\lambda 9531$} & Reference\\
 & Old & New  & Old & New & \\
\hline

Orion Nebula Cut~1 & $19.7 \pm 1.0$ & $32.2\pm 1.9$& $82.8  \pm 5.0$ & $84.7 \pm 5.9$ & \multirow{4}{*}{\citet{mendez2021}}\\

Orion Nebula Cut~2 & $20.8 \pm 1.3$ &  $32.5 \pm 2.3$& $91.1\pm 5.5$&$91.4 \pm 6.4 $\\

Orion Nebula Cut~3 & $21.8\pm 1.3$ & $31.8\pm 1.9$& $91.1\pm 6.4$&$91.4 \pm 6.4$\\

Orion Nebula Cut~4 &$22.8 \pm 1.4$ & $ 34.1 \pm 2.4$ & $82.5\pm 4.9$ & $85.2\pm6.0$\\

Orion Nebula + NIL Cut~1 & $25.4\pm 1.0$ & $28.6\pm 1.4$ & $ 69.3 \pm 2.8$ & $69.3 \pm 3.5$ & \multirow{2}{*}{\citet{mendez2021-2}}\\

Orion Nebula Cut~2 & $24.8\pm 1.2$ & $30.7\pm 1.8$ & $74.6\pm 4.5$ & $74.6\pm 5.2$ \\

HH~529~II & \multicolumn{2}{c}{$40.1 \pm 2.4$} &  \multicolumn{2}{c}{$98.5 \pm 6.9$}&\multirow{2}{*}{\citet{mendez2021}}\\

HH~529~III & \multicolumn{2}{c}{$41.1 \pm 3.3$} &  \multicolumn{2}{c}{$100.4 \pm 9.0$}&\\

HH~204 & \multicolumn{2}{c}{$36.5 \pm 1.5$} & \multicolumn{2}{c}{$87.2\pm 3.5$}\\

NIL & $26.5\pm 2.9$ & $20.7\pm2.5$& $ 60.6\pm 6.7$& $55.5\pm 6.6$& \citet{mendez2021-2}\\

\hline
\end{tabular}
\end{table*}

\begin{figure*}
\includegraphics[width=\textwidth]{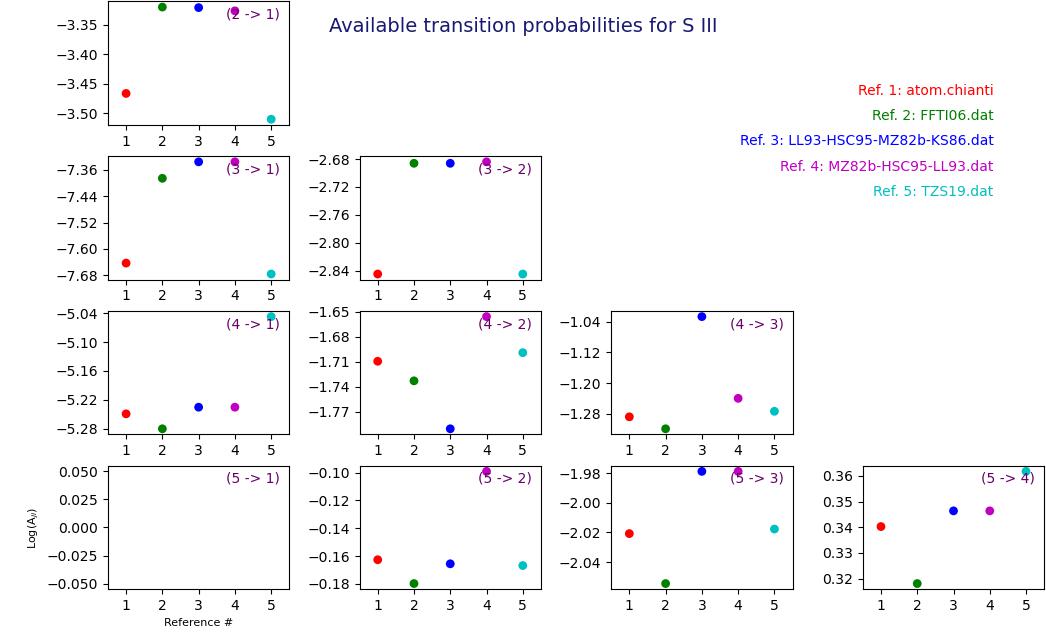}
\caption{Comparison between the different values of transition probabilities of [S\thinspace III] transitions given by the references studied in this paper.}
\label{fig:A_sii}
\end{figure*}

\begin{figure*}
\includegraphics[width=\textwidth]{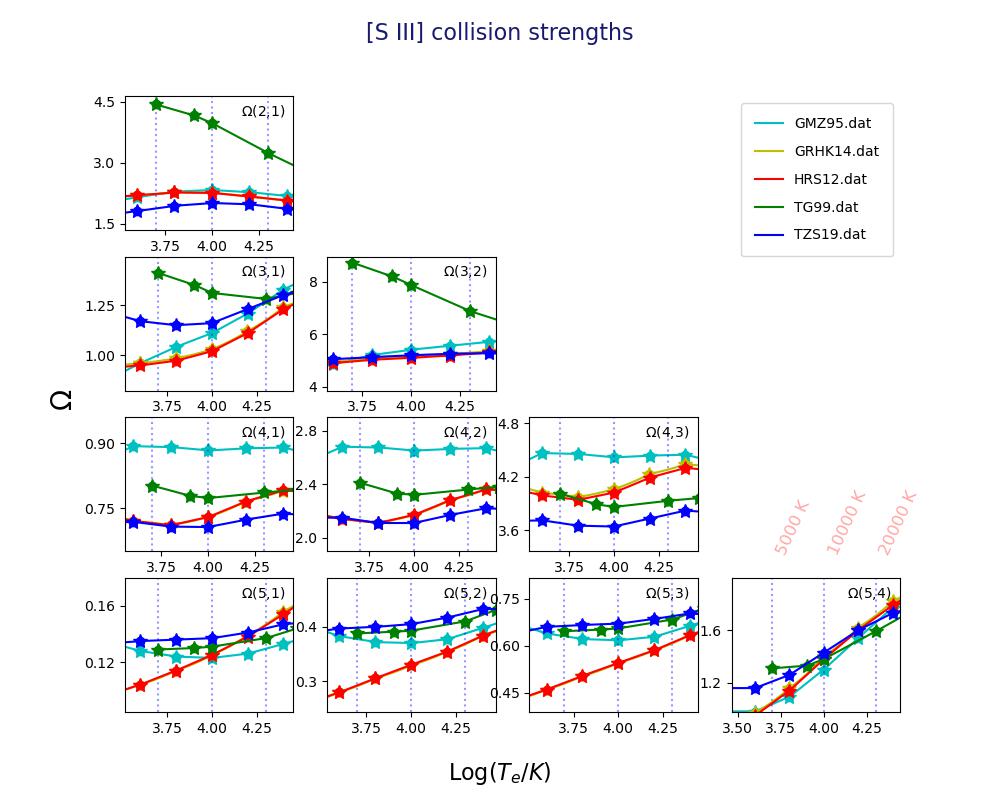}
\caption{Comparison between the different values of collision strengths of the [S\thinspace III] transitions given by the references studied in this paper.}
\label{fig:omega_siii_zoom }
\end{figure*}

\begin{table*}
\centering
\caption{$T_{\rm e} \text{ ([S\thinspace III])}$ and S$^{2+}$ abundance derived with each set of collision strengths and the transition probabilities from FFTI06. The last column has been determined increasing a 40 per cent the  $\Omega_{9531}$, $\Omega_{9069}$ and $\Omega_{8829}$ values of GRHK14.}
\label{tab:modified_data_omega}
\begin{adjustbox}{width=\textwidth}
\begin{tabular}{ccccccccccccc}
\hline
 &  GMZ95 & TG99 & HRS12 &  TZS19 & GRHK14 & Modified GRHK14\\
\hline

\multicolumn{7}{c}{$T_{\rm e} \text{ ([S\thinspace III])}$} \\ 

HH~514~jet base  & $8370^{+420} _{-450}$ & $7980^{+400} _{-530}$ & $8410^{+350} _{-390}$ & $7900^{+480} _{-170}$ & $8420^{+310} _{-400}$ & $8840^{+500} _{-560}$ \\

HH~514~knot & $8200^{+520} _{-510}$ & $7820^{+380} _{-490}$ & $8140^{+450} _{-600}$& $7750^{+310} _{-670}$ & $8230^{+430} _{-560}$ & $8980^{+570} _{-860}$\\

Nebular Cut~2 & $8850^{+270} _{-320}$ & $8270^{+280} _{-190}$ & $8650^{+270} _{-280}$ & - & $8630^{+330} _{-340}$ & $9860^{+400} _{-470}$\\
\hline

\multicolumn{7}{c}{12 + log(S$^{2+}$/H$^{+}$)} \\

HH~514~jet base & $7.35^{+0.07} _{-0.06}$ & $7.44^{+0.09} _{-0.07}$&$7.39 \pm 0.06 $& $7.47^{+0.08} _{-0.07}$& $7.39^{+0.07} _{-0.06}$& $7.25^{+0.08} _{-0.07}$\\

HH~514~knot & $7.30^{+0.09} _{-0.07}$ & $7.39^{+0.09} _{-0.07}$& $7.36^{+0.10} _{-0.08}$& $7.43^{+0.12} _{-0.09}$ & $7.35^{+0.10} _{-0.07}$&  $7.14^{+0.12} _{-0.09}$\\

Nebular Cut~2 &$6.76^{+0.05} _{-0.04}$ &$6.88^{+0.05} _{-0.04}$ &$6.86^{+0.05} _{-0.04}$ & -& $6.86 \pm 0.05 $ & $6.60^{+0.06} _{-0.05}$ \\

\hline
\end{tabular}
\end{adjustbox}
\end{table*}

\begin{table}
\caption{Intensities ratios of selected lines of  the spectrum of proplyd 170-337 after subtracting the emission of the Orion Nebula. c(H$\beta$)=$1.16\pm 0.14$, $F$(H$\beta$)=$2.54\times10^{-13}$(erg cm$^{-2}$ s$^{-1}$)} 
\label{tab:fluxes_proplyd}
\begin{tabular}{ccccccccc} 
\hline
$\lambda_0$ ( \AA ) & Ion & F$\left( \lambda \right)$/F$\left( \mbox{H}\beta \right)$ & I$\left( \lambda \right)$/I$\left( \mbox{H}\beta \right)$ & Err \% \\
\hline
4068.60 & $\mbox{[S}\thinspace \mbox{II]}$ & 3.42 & 4.72 & 5  \\
4076.35 & $\mbox{[S}\thinspace \mbox{II]}$ & 0.95 & 1.31 & 5  \\
4363.21 & $\mbox{[O}\thinspace \mbox{III]}$ & 2.66 & 3.31 & 4 \\
5754.64 & $\mbox{[N}\thinspace \mbox{II]}$ & 6.7 & 4.82 & 6 &  \\
5006.84 & $\mbox{[O}\thinspace \mbox{III]}$ & 276 & 260 & 2 \\
6312.10 & $\mbox{[S}\thinspace \mbox{III]}$ & 3.65 & 2.20 & 8 &  \\
6583.46 & $\mbox{[N}\thinspace \mbox{II]}$ & 44.4 & 24.4 & 8 &  \\
9530.60 & $\mbox{[S}\thinspace \mbox{III]}$ & 180 & 43 & 26 \\
\hline
\end{tabular}
\end{table}

\bsp	
\label{lastpage}
\end{document}